\documentclass[11pt,a4paper]{article}
\pdfoutput=1 

\usepackage{jheppub}
\usepackage{color}
\usepackage{multirow}
\newcommand{\Nvecs}{N_{\text{vecs}}}

\title{Excited and exotic charmonium spectroscopy from lattice QCD}

\author[a,1]{Liuming~Liu\note{liuming@maths.tcd.ie}}
\author[a]{Graham~Moir}
\author[a,2]{Michael~Peardon\note{mjp@maths.tcd.ie}}
\author[a,3]{Sin\'ead~M.~Ryan\note{ryan@maths.tcd.ie}}
\author[a,4]{Christopher~E.~Thomas\note{thomasc@maths.tcd.ie}}
\author[a]{Pol~Vilaseca}

\affiliation[a]{\vspace{0.1cm} School of Mathematics, Trinity College, Dublin~2, Ireland}

\author[b,c]{\\ \vspace{0.2cm} Jozef~J.~Dudek}
\author[b]{Robert~G.~Edwards}
\author[b]{B\'alint~Jo\'o}
\author[b]{David~G.~Richards}

\affiliation[b]{Jefferson Laboratory, 12000 Jefferson Avenue,  Newport News, VA 23606, USA}
\affiliation[c]{Department of Physics, Old Dominion University, Norfolk, VA 23529, USA}

\affiliation{\vspace{0.3cm} {\rmfamily \normalsize (for the Hadron Spectrum Collaboration)}}

\preprint{\begin{tabular}{r}TCDMATH 12-04\\JLAB-THY-12-1510\end{tabular}}

\arxivnumber{1204.5425}

\abstract{We present a spectrum of highly excited charmonium mesons up to around $4.5$~GeV calculated using dynamical lattice QCD.  Employing novel computational techniques and the variational method with a large basis of carefully constructed operators, we extract and reliably identify the continuum spin of an extensive set of excited states, states with exotic quantum numbers ($0^{+-}$, $1^{-+}$, $2^{+-}$) and states with high spin.  
Calculations are performed on two lattice volumes with pion mass $\approx 400~\text{MeV}$ and the mass determinations have high statistical precision even for excited states.
We discuss the results in light of experimental observations, identify the lightest `supermultiplet' of hybrid mesons and comment on the phenomenological implications of the spectrum of exotic mesons.}

\begin{document}
\maketitle

\section{Introduction}
\label{sec:intro}

There has been a resurgence of interest in charmonium spectroscopy over the 
last decade driven by experimental observations 
at the $B$-factories, CLEO-c, the Tevatron and BES of rather narrow states close to or 
above open-charm thresholds.
Some had been predicted but hitherto 
unobserved whereas many were unexpected, for example the enigmatic $X(3872)$, 
$Y(4260)$ and other ``X,Y,Z''s.
Prior to these discoveries, the relatively small number of observed 
mesons all fitted into the simple pattern predicted by quark models. The
new states have spurred theoretical discussion as to their nature with 
suggestions including conventional quark-antiquark states appearing at 
unexpected masses, hybrid mesons in which the gluonic field is excited, 
molecular mesons and tetraquarks.  To date, however, there has been no 
observation of a state in the charmonium region with manifestly exotic $J^{PC}$ 
($J,P$ and $C$ are respectively spin, parity and charge-conjugation quantum numbers); 
an observation of such an exotic would be a smoking gun for physics beyond 
quark potential models.  Experimental interest continues with BESIII, 
experiments at the LHC, and the planned PANDA experiment at GSI/FAIR.  An 
extensive review of the experimental and theoretical situation is given in 
Ref.~\cite{Brambilla:2010cs}.

From this perspective, the charmonium system is a particularly interesting 
probe of the strong regime of QCD.  Being reasonably non-relativistic it is 
well described below the open-charm ($D\bar{D}$) threshold by modelling it
as a quark and anti-quark bound together in a confining potential. It has
also been studied using effective field theory approaches and in the framework
of QCD sum rules.  No real consensus has emerged about the nature of many of 
the new states.

A vital part of understanding and testing QCD is to make precise predictions of the hadron spectrum it implies, as opposed to the spectrum in a QCD-inspired approach, and to test these predictions against high-quality experimental data.  The charmonium system provides an important arena for this and, in particular, discerning the puzzling states within QCD is a crucial aspect of understanding the strong interaction.  Lattice QCD provides a method for performing such ab-initio calculations within QCD: the theory is discretised on a finite lattice and then energies and other properties are extracted from numerical computations of Euclidean-time correlation functions.

The precision calculation of the masses of the lower-lying states has long been an important benchmark of lattice calculations; examples of recent studies of the lightest charmonium mesons are given Refs.~\cite{Follana:2006rc,Burch:2009az,Namekawa:2011cx,Mohler:2011ke} and reviewed in Ref.~\cite{DeTar:2011nn}.  More recently, progress has been made in using lattice QCD to study higher-lying states.  A pioneering quenched calculation of the excited charmonium spectrum was presented in Ref.~\cite{Dudek:2007wv} and there have since been some dynamical studies~\cite{Bali:2011dc,Bali:2011rd}; we comment on these previous studies later when we discuss the results of this work.

In lattice calculations, the theory is discretised on a finite hypercubic grid and the full rotational symmetry of the continuum is reduced to the symmetry group of a cube.  States at rest are then not classified according to spin, $J$, but by the irreducible representations, \emph{irreps}, of this reduced symmetry group; the various components of high spin states are distributed across several irreps.  The Hadron Spectrum Collaboration has overcome the challenges this reduced symmetry presents by using a large basis of carefully constructed operators with a combination of techniques and dynamical anisotropic lattices.  This methodology has been shown to be very effective at extracting extensive spectra of light isovector and isoscalar mesons~\cite{Dudek:2009qf,Dudek:2010wm,Dudek:2011tt} and baryons~\cite{Bulava:2010yg,Edwards:2011jj,Dudek:2012ag}, including highly excited states, states with high spin and those with exotic quantum numbers, and in reliably determining the continuum $J^{PC}$ of the extracted states.  

In this article we present a dynamical lattice calculation of the excited spectrum of charmonium mesons, the first application to charmonium of these techniques.  The results presented here go beyond any previous dynamical calculation of the excited charmonium spectrum.  We will show how we can reliably extract an unprecedented number of states, including the spectrum of exotics below around 4.5~GeV, and identify their continuum quantum numbers.  In particular, identification of the $J^{PC}$ will be vital in order to compare with experiment and so to disentangle the nature of the many puzzling resonances.  Preliminary results from this work have been presented in Ref.~\cite{Liu:2011rn}.

\section{The anisotropic lattice action}
\label{sec:lattices}

This study aims to determine the spectrum of charmonium states, including excitations and 
any states with an intrinsic gluonic component. 
In Euclidean space-time, this spectrum can be computed by
observing the behaviour of correlation functions of appropriately constructed
operators, which fall exponentially as the time separation is increased with 
rate in proportion to the energies of the eigenstates of the 
Hamiltonian. Since the charmonium spectrum lies above 3 GeV, these correlation 
functions decrease rapidly at large Euclidean time separations. Correlation 
functions of creation operators which excite high-lying states also suffer 
from substantial statistical variance in Monte Carlo determinations. The 
anisotropic lattice provides a very useful framework to ameliorate these 
technical problems, as it enables the correlator to be determined with a finer
temporal resolution at a manageable computational cost. 

The anisotropic lattice offers a further advantage when studying charmonium
physics.  The charm quark mass, $m_c$, has proved problematic in simulations in 
the past. It is too light for non-relativistic approaches
to be valid whilst it is also large enough that $am_c \ll 1$ only 
for the finest isotropic ensembles currently available. A number of successful
approaches have been developed, including the Fermilab and HISQ
actions~\cite{Follana:2006rc,ElKhadra:1996mp}.  
For this study, we will use a discretisation in which the spatial lattice 
spacing, $a_s$, and temporal lattice spacing, $a_t$, are related via $\xi=a_s/a_t\approx 3.5$. 
The fine temporal discretisation will ensure that we carry out our calculation with 
$a_tm_c \ll 1$ and the standard relativistic
formulation of fermion actions can be used to study states where the 
typical spatial momentum of a charm quark inside a hadron at rest is small.
This should be a reasonable approximation for the spectroscopy measurements we make in the current work. 
Nevertheless, in the simulations carried out here, $a_sm_c$ is also
less than unity.  We explore the size of $\mathcal{O}(a_s)$ errors in Section \ref{subsec:c_sw}.

\subsection{The lattice action}
\label{subsec:action}

For this study, we make use of the substantial ensembles of anisotropic 
gauge-field configurations generated by the Hadron Spectrum 
Collaboration~\cite{Edwards:2008ja,Lin:2008pr}. 
The gauge action is tree-level Symanzik-improved while in the fermion sector 
we use the anisotropic Shekholeslami-Wohlert action with tree-level tadpole
improvement and three-dimensional stout-link smearing~\cite{Morningstar:2003gk} of gauge fields.  
The fields are drawn from the importance sampling distribution 
appropriate for studies with two dynamical light quarks and a dynamical 
strange quark ($N_f = 2+1$); more details are given in Refs.~\cite{Edwards:2008ja,Lin:2008pr}.  

The charm quark action is the same as that used for the light and strange 
quarks, except that the parameter describing the weighting of spatial and
temporal derivatives in the lattice action is chosen to give a relativistic 
dispersion relation for low-momentum mesons containing charm quarks. 
This non-perturbative determination~\cite{Edwards:2008ja} is described 
in more detail in Section~\ref{subsec:aniso}.  
The charm quark fields are not treated dynamically in the configuration 
generation stage and so the quenching of these fields introduces a systematic 
uncertainty in our analysis. We expect the effects on the spectrum of this 
non-unitary treatment of charm quarks to be fairly small. 
The bare mass parameter for the valence charm quarks is determined by 
ensuring the ratio of the $\eta_c$ meson and the $\Omega$ baryon masses takes
its experimental value.

To quote energies and lengths in physical units we follow Ref.~\cite{Lin:2008pr} and consider the ratio of the mass of the $\Omega$ baryon measured on these ensembles, $a_t m_{\Omega} = 0.2951(22)$~\cite{Edwards:2011jj}, to the experimental mass, $m_{\Omega} = 1672.45(29)$~MeV~\cite{PDG2011}.
Setting the scale in this way, we find the parameters in the lattice action 
used in this study correspond to a spatial lattice spacing of $a_s \sim 0.12$~fm 
and a temporal lattice spacing approximately 3.5 times smaller, $a_t^{-1} \sim 5.7$ GeV.
The lattices employed here have $m_{\pi} \approx 400~\text{MeV}$.
Two volumes are used, 
$(L/a_s)^3 \times (T/a_t) = 16^3 \times 128$ and $24^3 \times 128$, 
corresponding to spatial extents of $\sim 1.9$ fm and $\sim 2.9$ fm 
respectively.  The ensembles used in this study are summarised in 
Table~\ref{tab:latt_details} and full details of the lattice actions 
and parameters are given in Refs.~\cite{Edwards:2008ja,Lin:2008pr}.

\begin{table}[t]
\begin{center}
\begin{tabular}{ccccc}
Volume & $m_\pi/$MeV & $N_{\rm cfgs}$ & $N_{\rm tsrcs}$ & $N_{\rm vecs}$ \\
\hline
$16^3\times 128$ & 396 &  96 & 128 & 64  \\
$24^3\times 128$ & 396 & 552 &  32 & 162
\end{tabular}
\caption{The gauge-field ensembles used in this work.  $N_{\rm cfgs}$ and 
$N_{\rm tsrcs}$ are respectively the number of gauge-field configurations and 
time-sources per configuration used; $N_{\rm vecs}$ refers to
the number of eigenvectors used in the distillation method~\cite{Peardon:2009gh}. }
\label{tab:latt_details}
\end{center}
\end{table}

\subsection{Determining the action parameters for the charm quark}
\label{subsec:aniso}

On the anisotropic lattice, the weights of any action terms that represent
temporal and spatial derivatives must be determined in order that a physical
observable takes its experimental value when calculated in the simulation. This is of
particular importance for dynamical simulations to ensure the correct continuum
limit~\cite{Morrin:2006tf}. For these ensembles where the target ratio of 
scales is $\xi=3.5$, this determination for the gauge fields along 
with the dynamical light and strange quarks is presented in 
Ref.~\cite{Edwards:2008ja}.  This ratio has been measured in the light meson
sector from the pion dispersion relation, giving $\xi_\pi=3.444(6)$~\cite{Dudek:2012gj}. 

As part of this study, we determined the dispersion relation for the $\eta_c$ 
meson on the $16^3$ ensemble.
The dispersion relation for meson $A$ is 
\begin{equation}
  (a_t E_A)^2 = (a_t m_A)^2 + \left(\frac{1}{\xi_A}\right)^2 (a_s p)^2, 
\end{equation}
and since all quark fields making up the meson obey periodic boundary 
conditions for the finite spatial extent of the lattice, the momentum values 
are quantized, so 
$a_s \vec{p} = \frac{2\pi}{L}(n_x,n_y,n_z)$ with 
$n_i\in \{0,1,\dots,L/a_s-1\}$. 
Since we plan to follow up this work with a study of the scattering of mesons including 
charm quarks and hope to investigate the physics of states above the 
open-charm threshold, it is important to check that the dispersion relation is correct 
and that $\xi$ is consistent with that in the light quark sector.

\begin{figure}[t]
\centering
\includegraphics*[width=0.48\textwidth]{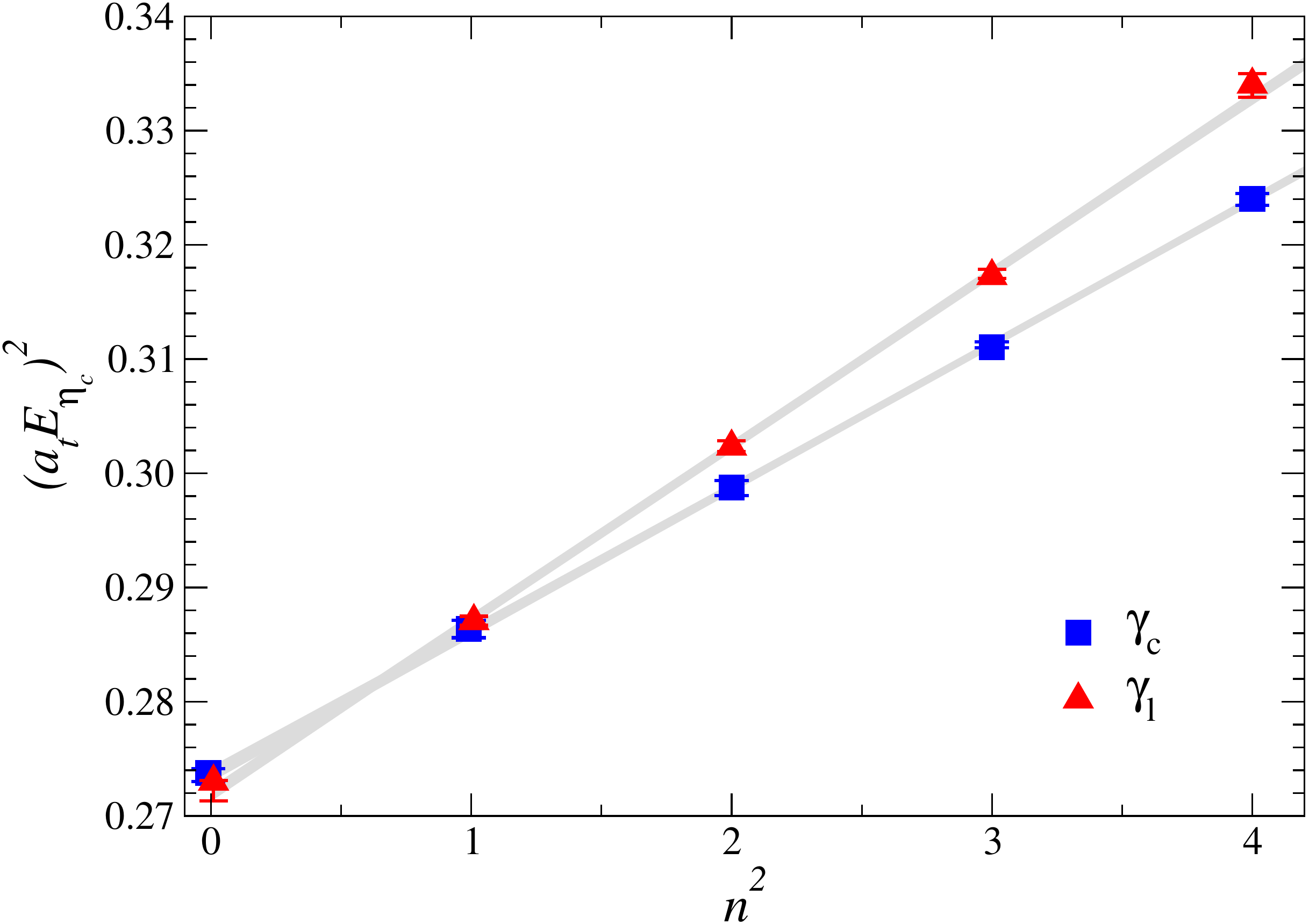} \hfill
\includegraphics*[width=0.48\textwidth]{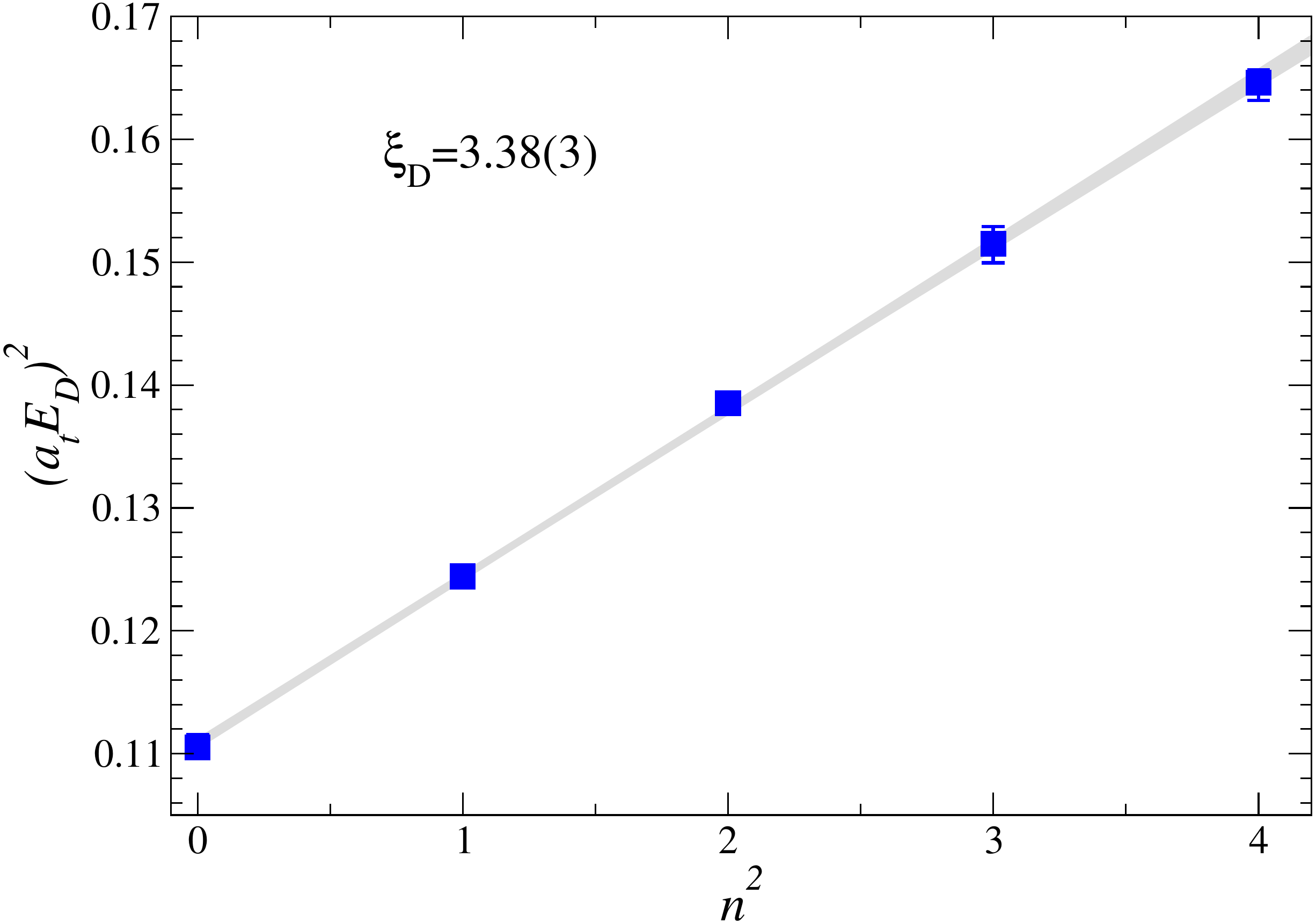}
\caption{Left panel: the dispersion relation of the $\eta_c$ meson measured on 96 configurations
for two choices of the bare parameter, $\gamma_f$, in the charm quark action.
As described in the text, $\gamma_l$ is the parameter in the light
sea-quark action and $\gamma_c$ is determined by ensuring the correct dispersion
relation for the $\eta_c$ meson.  Right panel: the dispersion relation for the
heavy-light $D$ meson measured on 39 configurations with
$\gamma_f = \gamma_c$ in the charm quark action.  Both on $16^3$ volumes. } 
\label{fig:dispersion}
\end{figure}

In the left panel of Fig.~\ref{fig:dispersion}, 
the dependence of the $\eta_c$ meson energy with increasing quantized 
momenta, labelled by $n^2=n_x^2+n_y^2+n_z^2$, is plotted, along with the best
straight-line fits. 
The triangular symbols show the dispersion relation when the bare parameter, $\gamma_f$, 
in the charm quark action is the same as that used for the light sea quarks, 
$\gamma_f = \gamma_l=3.4$~\cite{Edwards:2008ja}.
From this data, the measured ratio of scales is $\xi_{\eta_c} = 3.18(2)$, 
a discrepancy from the target value of about $9\%$. 

To correct for this discrepancy,  
simulations were performed to determine the charm-quark action 
parameters that recover a consistent dispersion relation.  Increasing the bare parameter 
to $\gamma_f = \gamma_c=3.988$, gives
the second dispersion relation denoted by squares in the left panel of Fig.~\ref{fig:dispersion}. 
The fit now yields a measured value of $\xi_{\eta_c}  = 3.50(2)$, consistent with 
the target value.
Note that the statistical uncertainty on this determination is below the percent-level. 
We employ this $\gamma_f = \gamma_c$ in the charm quark action in all subsequent computations.

As mentioned above, our future plans include a study of scattering of 
$D$ mesons, so it is important to check whether the self-consistent action 
parameters presented above also result in a relativistic dispersion relation 
for the $D$ meson. The light quark action parameters are left unchanged from 
those in the sea. This dispersion relation is presented in the right panel of 
Fig.~\ref{fig:dispersion} and the resulting slope yields a 
determination of the ratio of scales as $\xi_{D}=3.38(3)$, just $3\%$ 
different from the target value. The $D$ meson mass extracted from this
analysis is $M_{D}=1893(1)$ MeV, but note that the simulated light quark mass 
is greater than its physical value, $m_{\pi} \approx 400$~MeV.

We also note that the data in each of the figures agree 
with the relativistic dispersion relation up to $\vec{n}=(2,0,0)$. All 
five momenta are used in the linear fits which have $\chi^2/N_{\rm dof}\sim 1$.

\section{Operator construction and correlator analysis}
\label{sec:methods}

In lattice calculations, meson masses are extracted from two-point correlation functions,
\begin{equation}
C_{ij}(t)= \langle 0|{\cal O}_i(t){\cal O}^{\dagger}_j(0)|0\rangle ~,
\end{equation}
where the interpolating operators ${\cal O}^{\dagger}_j(0)$ create the state of interest at $t=0$ and 
${\cal O}_i(t)$ annihilate the state at a later Euclidean time $t$.  Inserting a complete set of 
eigenstates of the Hamiltonian we obtain the spectral decomposition,
\begin{equation}
C_{ij}(t) = \sum_\mathfrak{n} \frac{Z_i^{\mathfrak{n}*} Z_j^{\mathfrak{n}}}{2E_\mathfrak{n}} e^{-E_\mathfrak{n}t}  ~,
\end{equation}
where the vacuum-state matrix elements,
\begin{equation}
\label{equ:Z}
Z^{\mathfrak{n}}_i \equiv \langle \mathfrak{n} | \mathcal{O}^{\dagger}_i | 0 \rangle  ~,
\end{equation}
are referred to as \emph{overlaps}, and the sum is over a discrete set of states because the calculation is performed in a finite volume.  It is essential that the operators overlap well with the states under consideration.  A well-established way to achieve this is to apply a smearing to the quark fields in the operators to emphasise the relevant low-energy modes.

We use the \emph{distillation} technique, described in Ref.~\cite{Peardon:2009gh}, which implements a smearing and provides an efficient method to calculate correlation functions with large bases of operators.  Furthermore, it allows us to project onto definite momentum at both the source and the sink with reasonable computational cost.  Distillation defines a smearing function,
\begin{equation}
\Box_{xy}(t) = \sum_{k=1}^{\Nvecs} F(\lambda^{(k)}) \; \xi_x^{(k)}(t) \xi_y^{(k)\dag}(t) ~,
\label{equ:distillation}
\end{equation}
where in this work $\lambda^{(k)}$ and $\xi^{(k)}$ are the eigenvalues and eigenvectors of the gauge-covariant lattice Laplacian, $-\nabla^2$, evaluated on the background of the spatial gauge-fields of time-slice $t$ and sorted by eigenvalue.  Here $F(\lambda^{(k)})$ is some smearing profile and in this work we set $F = 1$.  The distillation operator, $\Box$, is a projection operator into the subspace spanned by these eigenmodes and so $\Box^2 = \Box$.  Smeared quark fields are constructed by applying this distillation operator to each quark field appearing in the interpolating operators, $\tilde{\psi} = \Box \psi$.  As described in Ref.~\cite{Peardon:2009gh}, the outer product structure of $\Box$ allows a correlation function to be factorised.  The computational cost is manageable because inversions are performed in the smaller space of distillation vectors and spin, with dimensionality $\Nvecs N_{\sigma}$, where $N_{\sigma} = 4$ is the number of components of a lattice Dirac spinor.  Further, this factorisation allows the efficient computation of correlation functions with a large basis of operators, as well as three-point functions and multi-hadron two-point correlation functions.  All the quark fields referred to subsequently are smeared using the distillation operator and we denote them by $\psi$.

The simplest interpolating operators are smeared local fermion bilinears of the form ${\cal O}=\bar{\psi}(x)\Gamma\psi(x)$, where $\Gamma$ is any product of Dirac gamma matrices and for clarity we have suppressed spin, flavour and colour indices.  However, such operators do not give access to $J \ge 2$ or even to the complete set of $J^{PC}$ quantum numbers for $J \le 1$.  In addition, as will be discussed later, we require a large basis of operators in each quantum number channel in order to reliably and precisely extract the energies of excited states. 

To give us a large basis of operators with a range of spatial structures in each channel, we use the derivative-based construction for operators described in Refs.~\cite{Dudek:2009qf,Dudek:2010wm}: gauge-covariant spatial derivatives are combined with a gamma matrix within a fermion bilinear.  The operator is of the general form
\begin{equation}
\bar{\psi}\Gamma \overleftrightarrow{D}_i \overleftrightarrow{D}_j \cdots\psi ~,
\end{equation}
where $\overleftrightarrow{D} \equiv \overleftarrow{D} - \overrightarrow{D}$ is a lattice-discretised gauge-covariant derivative.  As discussed in Ref.~\cite{Dudek:2010wm}, operators with definite $J^{PC}$ can be constructed using these ``forward-backward'' derivatives.  The derivatives and vector-like gamma matrices are first expressed in the circular basis so that they transform as $J=1$.  For creation operators the basis is
\begin{equation}
\overleftrightarrow{D}_{m=\pm1} = \mp\frac{i}{\sqrt{2}}(\overleftrightarrow{D}_x \pm i \overleftrightarrow{D}_y), \quad 
\overleftrightarrow{D}_{m=0} = i \overleftrightarrow{D}_z  ~. 
\end{equation} 
Using the standard $SO(3)$ Clebsch-Gordan coefficients, we can combine any number of derivatives with gamma matrices to construct operators with the desired quantum numbers, $\mathcal{O}^{J,M}$, where $M$ is the $J_z$ component.  The choice of $\Gamma$ and the way in which the derivatives have been coupled determine the parity, $P$, and charge conjugation, $C$, quantum numbers.  The notation we use follows that of Ref.~\cite{Dudek:2010wm}: an operator $(\Gamma \times D^{[N]}_{J_D})^{J}$ contains a gamma matrix, $\Gamma$, with the naming scheme given in Table \ref{table:GammaMatrix}, and $N$ derivatives coupled to spin $J_D$, overall coupled to spin $J$.

\begin{table}[t]
\begin{center}
\begin{tabular}{l l | cccccccc}
         & &$a_0$ &$\pi$ &$\pi_2$ &$b_0$ &$\rho$ &$\rho_2$ &$a_1$ &$b_1$ \\
\hline
$\Gamma$ & &1 &$\gamma_5$ &$\gamma_0 \gamma_5$ &$\gamma_0$ &$\gamma_i$ &$\gamma_0 \gamma_i$ &$\gamma_5 \gamma_i$ & $\gamma_0 \gamma_5 \gamma_i$ 
\end{tabular}
\caption{Gamma matrix naming scheme.}
\label{table:GammaMatrix}
\end{center}
\end{table}

\begin{table}[t]
\begin{center}
\begin{tabular}{ccl}
$J$ & & $\Lambda(\text{dim})$ \\
\hline
$0$ & & $A_1(1)$ \\
$1$ & & $T_1(3)$ \\
$2$ & & $T_2(3) \oplus E(2)$\\
$3$ & & $T_1(3) \oplus T_2(3) \oplus A_2(1)$\\
$4$ & & $A_1(1) \oplus T_1(3) \oplus T_2(3) \oplus E(2)$
\end{tabular}  
\caption{Continuum spins, $J \le 4$, subduced into lattice irreps, $\Lambda(\text{dim})$, where dim is the dimension of the irrep.}
\label{Table:Subduce}
\end{center}
\end{table}

In lattice QCD calculations the full three-dimensional rotational symmetry of an infinite volume continuum is reduced to the symmetry group of a cube.  Instead of the infinite number of irreducible representations labelled by spin, $J \ge 0$, we instead have a finite number of lattice irreducible representations (\emph{irreps}); the five single-cover irreps for states at rest are $A_1$, $T_1$, $T_2$, $E$ and $A_2$.  Note that parity and any flavour quantum numbers such as charge conjugation are still good quantum numbers on the lattice.  The distribution of the various components, $M$, of a spin-$J$ state across the lattice irreps is known as \emph{subduction} and in Table \ref{Table:Subduce} we show the pattern of these subductions for $J \le 4$.  Because a state with $J>1$ is split across many lattice irreps and each lattice irrep contains an infinite tower of $J$, it is not straightforward to identify the $J$ of states extracted in lattice computations.  Refs.~\cite{Dudek:2009qf,Dudek:2010wm} presented a method to address this problem and we discuss this further in Section \ref{sec:spin}.  

\begin{figure}[t]
\begin{center}
\includegraphics[width=1.0\textwidth]{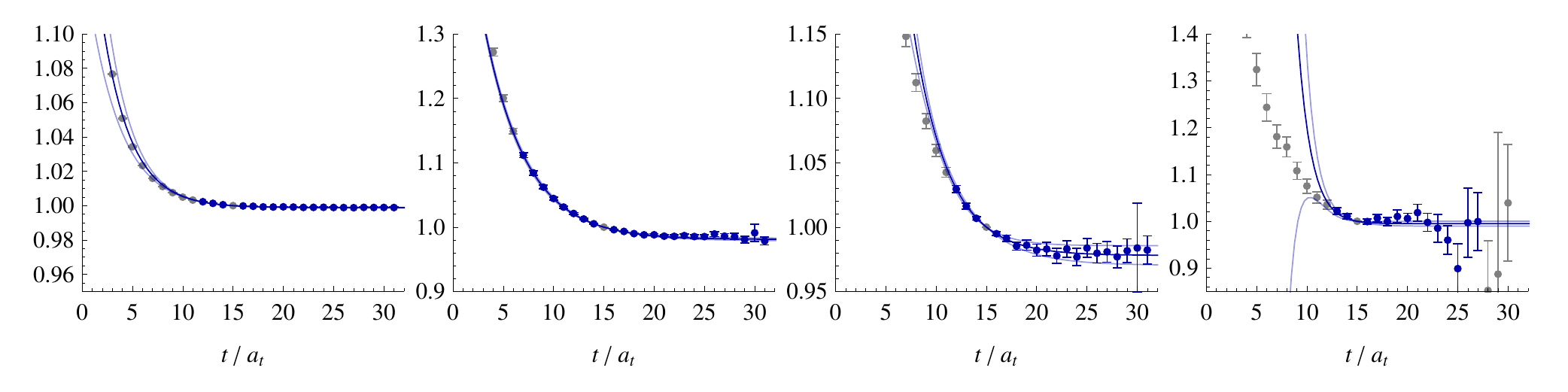}
\caption{Principal correlator fits according to Eq.~\eqref{equ:fitprincorr} for a selection of low-lying states in the $T_1^{--}$ irrep on the $24^3$ volume.  Plotted are $\lambda^{\mathfrak{n}}(t) \cdot e^{m_{\mathfrak{n}}(t-t_0)}$ data and the fits for $t_0 = 15$ showing the central values and one sigma statistical uncertainties; the grey points are not included in the fits.  From left to right the states have $a_t m = 0.53726(4)$, $0.6713(5)$, $0.676(1)$ and $0.753(2)$.  In all cases, for the $t_0$ used we obtain reasonable fits with $\chi^2/N_{\text{dof}} \sim 1$.}
\label{Fig:PrinCorrFits}
\end{center}
\end{figure}

Refs.~\cite{Dudek:2009qf,Dudek:2010wm} also discussed how to construct operators which transform in a definite lattice irrep, $\Lambda$, and row, $\lambda$, from the continuum operators, $\mathcal{O}^{J,M}$,
\begin{equation}
\mathcal{O}^{[J]}_{\Lambda,\lambda} = \sum_{M} \mathcal{S}^{\Lambda,\lambda}_{J,M} \mathcal{O}^{J,M} ~,
\end{equation}
where $\mathcal{S}^{\Lambda,\lambda}_{J,M}$ are subduction coefficients.  In this work we include operators built from all possible combinations of gamma matrices with up to three derivatives and then subduced into lattice irreps.  Table~\ref{table:ops_per_irrep} lists the numbers of operators in each irrep.

\begin{table}[t]
\begin{center}
\begin{tabular}{l l|cccc}
$\Lambda$ & & $\Lambda^{-+}$ & $\Lambda^{--}$ & $\Lambda^{++}$ & $\Lambda^{+-}$ \\
\hline
$A_1$     & & 12   &  6   & 13   &  5 \\
$A_2$     & & 4    &  6   & 5    &  5 \\
$T_1$     & & 18   & 26   & 22   & 22 \\
$T_2$     & & 18   & 18   & 22   & 14 \\
$E$       & & 14   & 12   & 17   &  9 \\
\end{tabular}
\caption{The number of operators used in each lattice irrep, $\Lambda^{PC}$.  All combinations of gamma matrices and up to three derivatives are included.}
\label{table:ops_per_irrep}
\end{center}
\end{table}

We use the variational method~\cite{Michael:1985ne,Luscher:1990ck} to extract spectral information from the two-point correlation functions with the particular implementation described in detail in Ref.~\cite{Dudek:2010wm}.  In brief, for each lattice irrep we form a matrix of correlators, $C_{ij}(t)$, where $i$ and $j$ label operators in our basis for that irrep.  The best extraction (in the variational sense) of the energies and overlaps then follows from solving a generalised eigenvalue problem,
\begin{equation}
C_{ij}(t) v^{\mathfrak{n}}_j = \lambda^{\mathfrak{n}}(t,t_0) C_{ij}(t_0) v^{\mathfrak{n}}_j ~,
\end{equation}
where an appropriate reference time-slice $t_0$ is chosen as described in Refs.~\cite{Dudek:2010wm,Dudek:2007wv}, the energies are determined by fitting the dependence of the eigenvalues, $\lambda^{\mathfrak{n}}$, called the \emph{principal correlators}, on $t - t_0$, and the overlaps are related to the eigenvectors, $v^{\mathfrak{n}}_j$. 

We fit the principal correlators using the fit form,
\begin{equation}
\lambda_\mathfrak{n}(t) = (1 - A_\mathfrak{n}) e^{-m_\mathfrak{n}(t-t_0)} + A_\mathfrak{n} e^{-m_\mathfrak{n}' (t-t_0)} ~,
\label{equ:fitprincorr}
\end{equation}
where the fit parameters are $m_\mathfrak{n}$, $m_\mathfrak{n}'$ and $A_\mathfrak{n}$.  This form allows for a second exponential and empirically we find that the contribution of this second exponential decreases rapidly as we increase $t_0$.  Some example fits to the principal correlators are shown in Fig.~\ref{Fig:PrinCorrFits}.  These are plotted with the dominant time-dependence due to state $\mathfrak{n}$ divided out and so we would see a horizontal line at 1 in the case where a single exponential dominates the fit.

The combination of the variational method, our operator constructions, the distillation technique, and these anisotropic lattice ensembles has been shown to be very effective in studies of excited light isovector~\cite{Dudek:2009qf,Dudek:2010wm} and isoscalar~\cite{Dudek:2011tt} mesons and baryons~\cite{Edwards:2011jj,Dudek:2012ag}.  In the following section we review our method for identifying the continuum spin of extracted states and present its first application to charmonium spectroscopy.

\section{Determining the spin of a state}
\label{sec:spin}

In principle, the spin of a single-hadron state can be determined by extracting the spectrum
at various lattices spacings and then extrapolating to the continuum limit.  
There, where full $SO(3)$ rotational symmetry is restored, energy degeneracies
between different lattice irreps should emerge.  For example, according to 
the pattern of subductions (Section \ref{sec:methods}), a spin-2 state would be 
expected to appear as degenerate energies within the $T_2$ and $E$ irreps. 
However, there are difficulties with this procedure.  First, this 
requires high precision calculations with successively finer lattice spacings and 
so with increasing computation cost; this is not practical with currently available 
resources.  Second, the continuum spectrum can exhibit degeneracies and this is
particularly true in the charmonium system where hyperfine splittings are small.
The question then arises as to how to identify, without infinite statistics, 
which degeneracies are due to the lattice discretisation and which are physical degeneracies.

\begin{figure}[t]
\begin{center}
\includegraphics[width=0.5\textwidth]{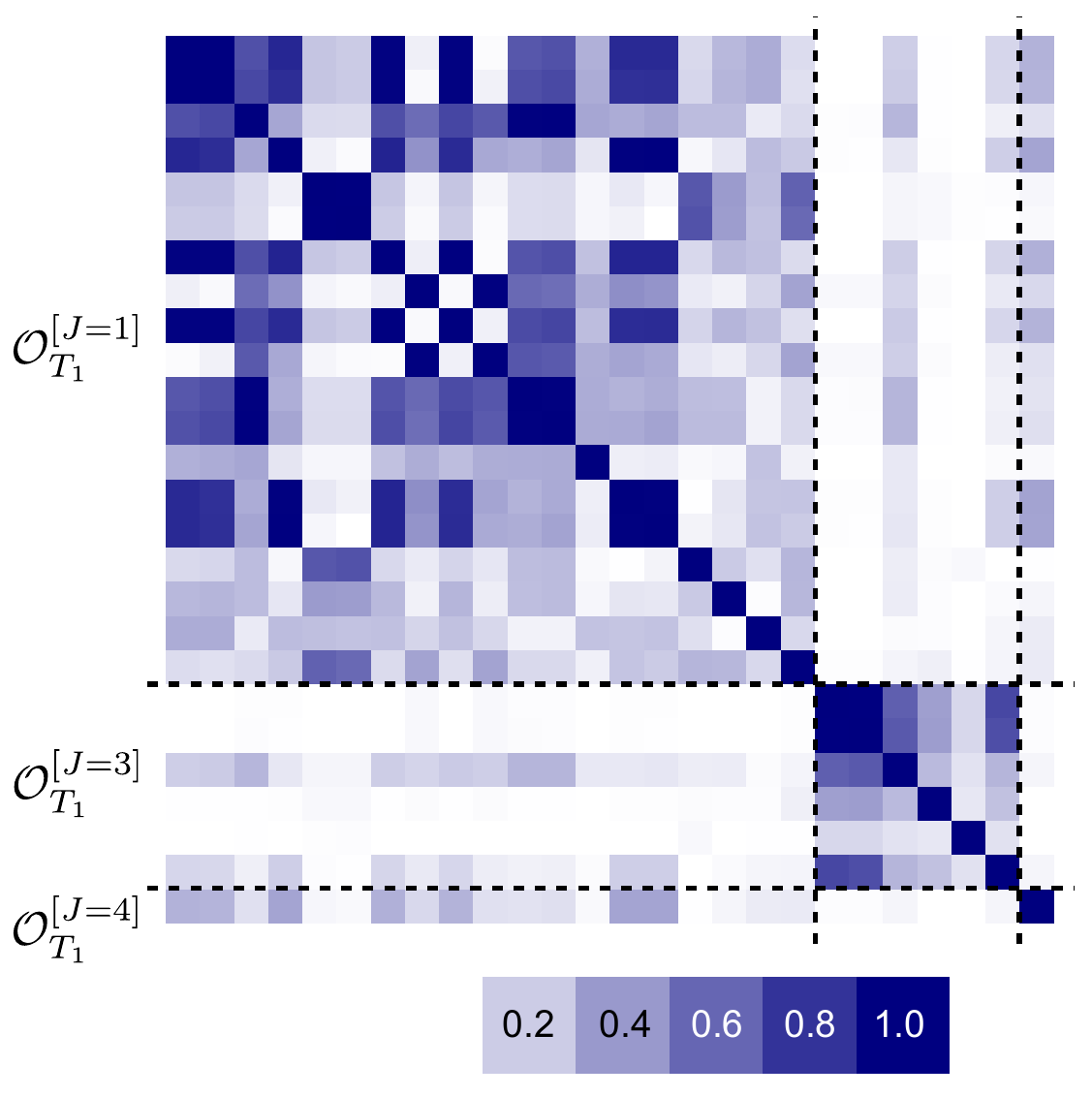}
\caption{The normalised correlation matrix, ($C_{ij}/\sqrt{C_{ii}C_{jj}}$) with $t/a_t=5$
in the $T_1^{--}$ irrep on the $24^3$ volume.  The operators are ordered such that those 
subduced from spin 1 appear first followed by spin 3 and then spin 4.
The correlation matrix is observed to be approximately block diagonal. }
\label{Fig:Corr_Matrix} 
\end{center}
\end{figure}

An alternative approach, proposed in Refs.~\cite{Dudek:2009qf,Dudek:2010wm}, is to consider 
the overlaps, Eq.~\eqref{equ:Z}, 
using this additional information to determine the spin of extracted states 
at a single lattice spacing.  Here we will only review the important points.
The subduced operators constructed in Section~\ref{sec:methods} respect the 
symmetries of the lattice.  However, we find that each operator, 
$\mathcal{O}^{[J]}_{\Lambda,\lambda}$ carries a ``memory" of the continuum spin, 
$J$, from which it is subduced.  If our lattice is reasonably close to restoring
continuum symmetry at the hadronic scale, we can expect that an operator subduced from spin $J$ will 
only overlap strongly onto those states of continuum spin $J$~\cite{Davoudi:2012ya}.

\begin{figure*}[t]
\begin{center}
\begin{tabular}{c}
\includegraphics[height=0.25\textwidth]{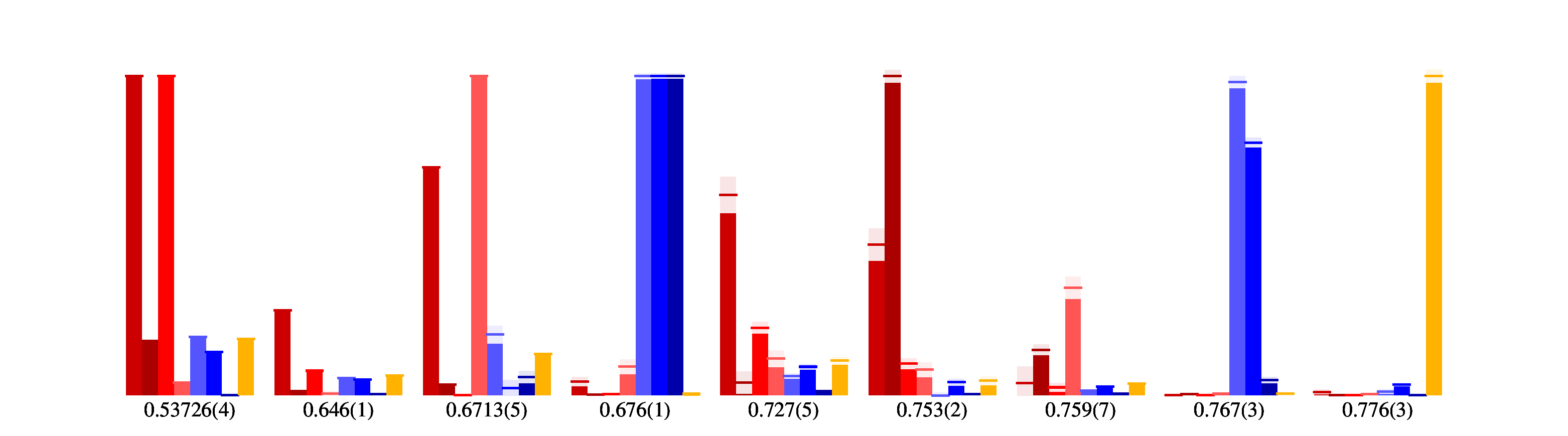} \\
$T_1^{--}$
\end{tabular}
\begin{tabular}{cccc}
\includegraphics[height=0.25\textwidth]{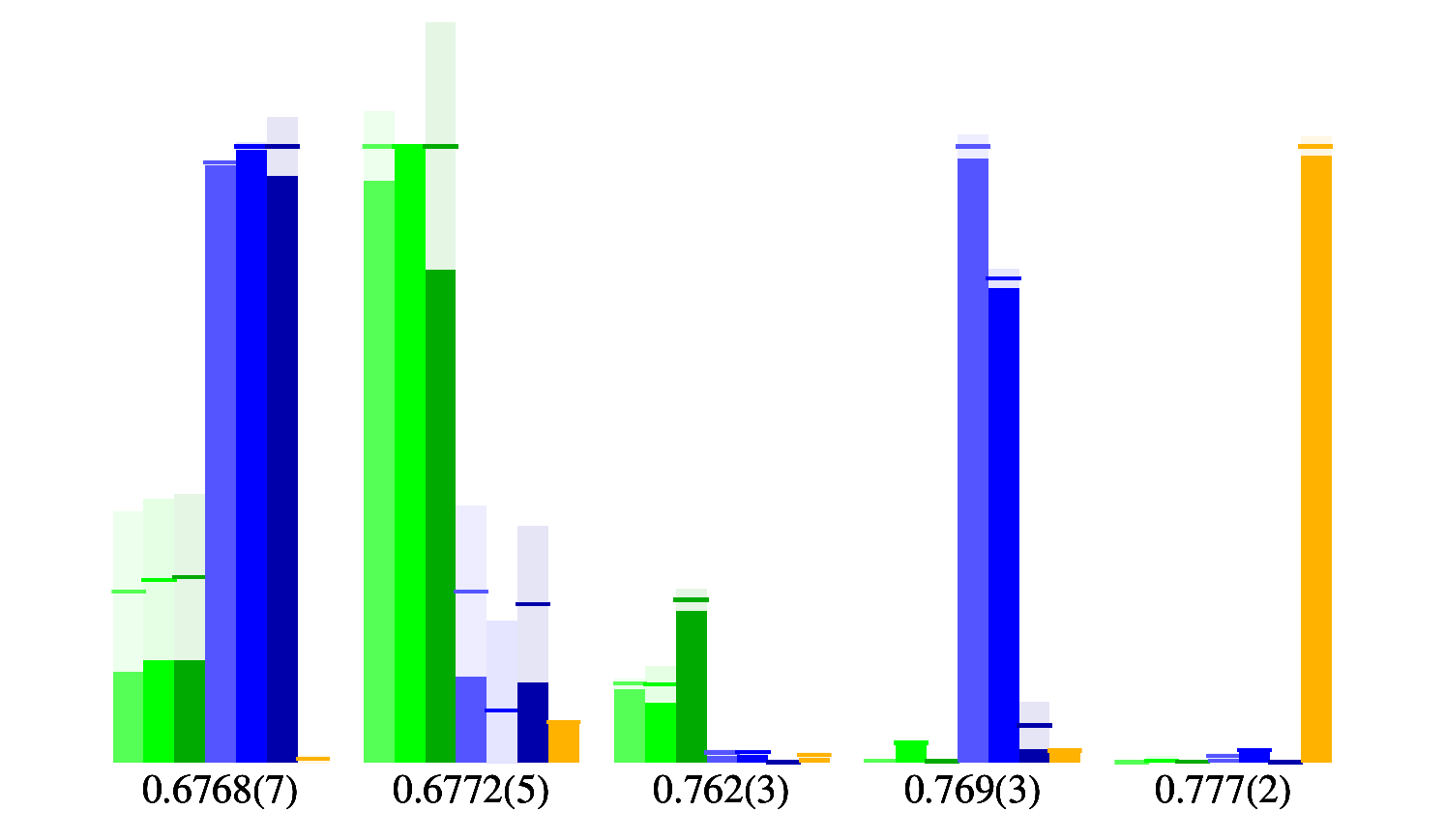} &
\includegraphics[height=0.25\textwidth]{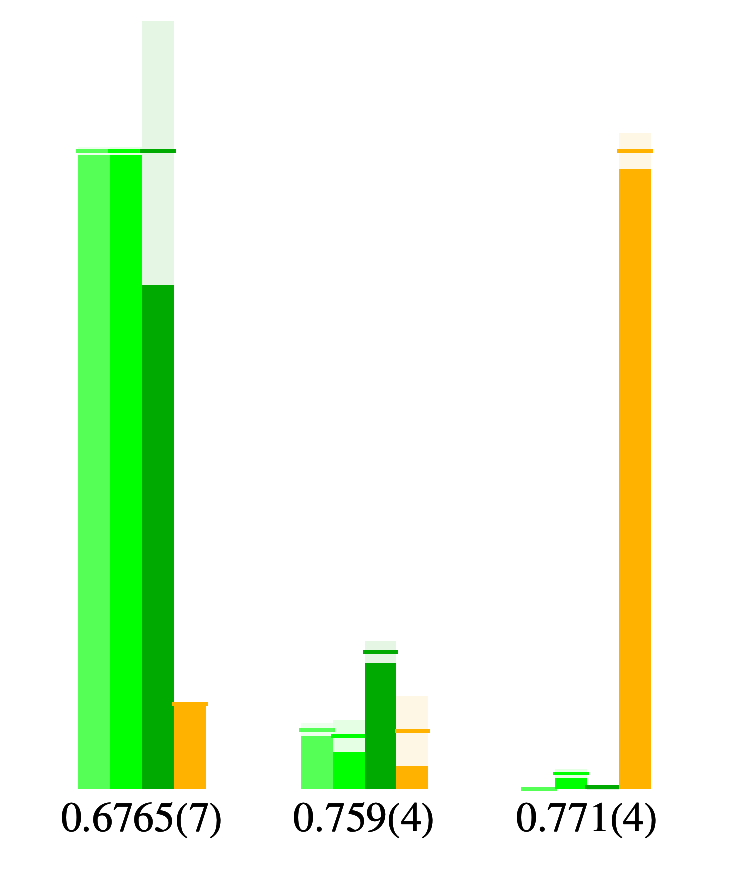} &
\includegraphics[height=0.25\textwidth]{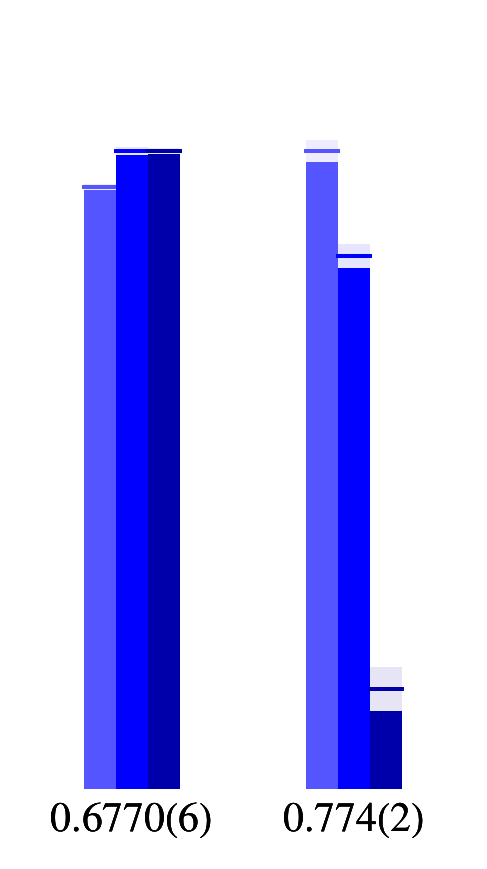} &
\includegraphics[height=0.25\textwidth]{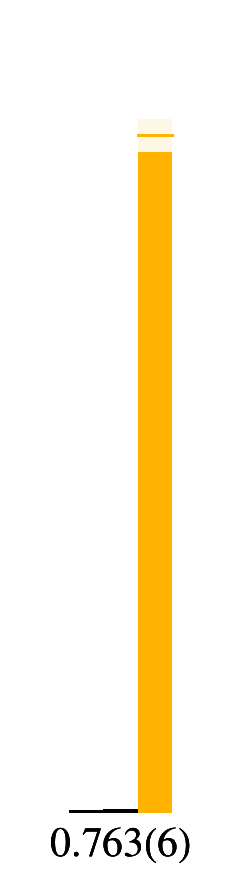}\\
$T_2^{--}$ & $E^{--}$ & $A_2^{--}$ & $A_1^{--}$
\end{tabular}
\includegraphics[width=0.5\textwidth]{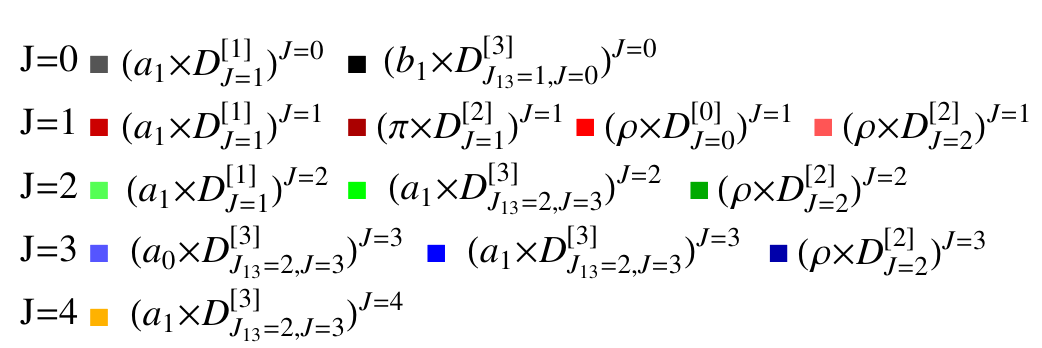}
\caption{The overlaps, $Z$, of a selection of operators onto some of the lower-lying states in each of the lattice irreps, $\Lambda^{--}$, on the $24^3$ volume; 
states are labelled by their mass in lattice units, $a_t m$.
The $Z$-values are normalized so that the largest value for that operator 
across all states is equal to 1.  The lighter area at the top of each bar 
represents the one sigma statistical uncertainty on either side of the mean.  
Different color shading represents different operators.}
\label{Fig:Jmm_histo}
\end{center}
\end{figure*}

This is apparent in both the correlation matrix and
the overlaps of individual states.  Fig.~\ref{Fig:Corr_Matrix} shows the 
correlation matrix for $T_1^{--}$ at time-slice 5 where the operators 
are ordered according to the spin from which they are subduced.  It can be
seen that the correlation matrix is approximately block diagonal.  
In Fig.~\ref{Fig:Jmm_histo} we show the overlaps of a selection of 
operators onto states in the irreps $\Lambda^{--}$.  The $J=0$ operators are colored black, 
$J=1$ are red, $J=2$ are green, $J=3$ are blue and $J=4$ are gold.  Every state has a clear 
preference to overlap only onto those operators subduced from a single continuum spin.  
For clarity we only show a subset of the operators but this pattern is 
observed for our full operator basis.

To be more quantitative, we can compare the overlaps obtained in different irreps. 
The continuum operators have definite spin so that 
$\langle J^\prime, M^\prime | \mathcal{O}^{J, M \dagger} | 0 \rangle = Z^{[J]} \delta_{J, J^\prime} \delta_{M, M^\prime}$ 
and therefore we have
$\langle J^\prime, M^\prime | \mathcal{O}^{[J] \dagger}_{\Lambda, \lambda} | 0 \rangle = S^{J, M^\prime}_{\Lambda, \lambda}Z^{[J]} \delta_{J, J^\prime}$.  
The value of $Z^{[J]}$ is common to different lattice irreps: 
 because the subduction coefficients form 
an orthogonal matrix, the spectral decomposition will contain terms proportional 
to $Z^{[J]*} Z^{[J]}$; these do not depend on the $\Lambda$ subduced into, up to
discretisation effects.  This suggests that we can identify the spin of a state 
by comparing the $Z$-values obtained independently in different irreps.  
For example, a $J=2$ state created by a $\mathcal{O}_{\Lambda}^{[J=2]}$ operator 
will have the same $Z$-value in each of the $T_2$ and $E$ irreps.  

\begin{figure*}[t]
\begin{center}
\begin{tabular}{ccc}
\includegraphics[height=0.36\textwidth]{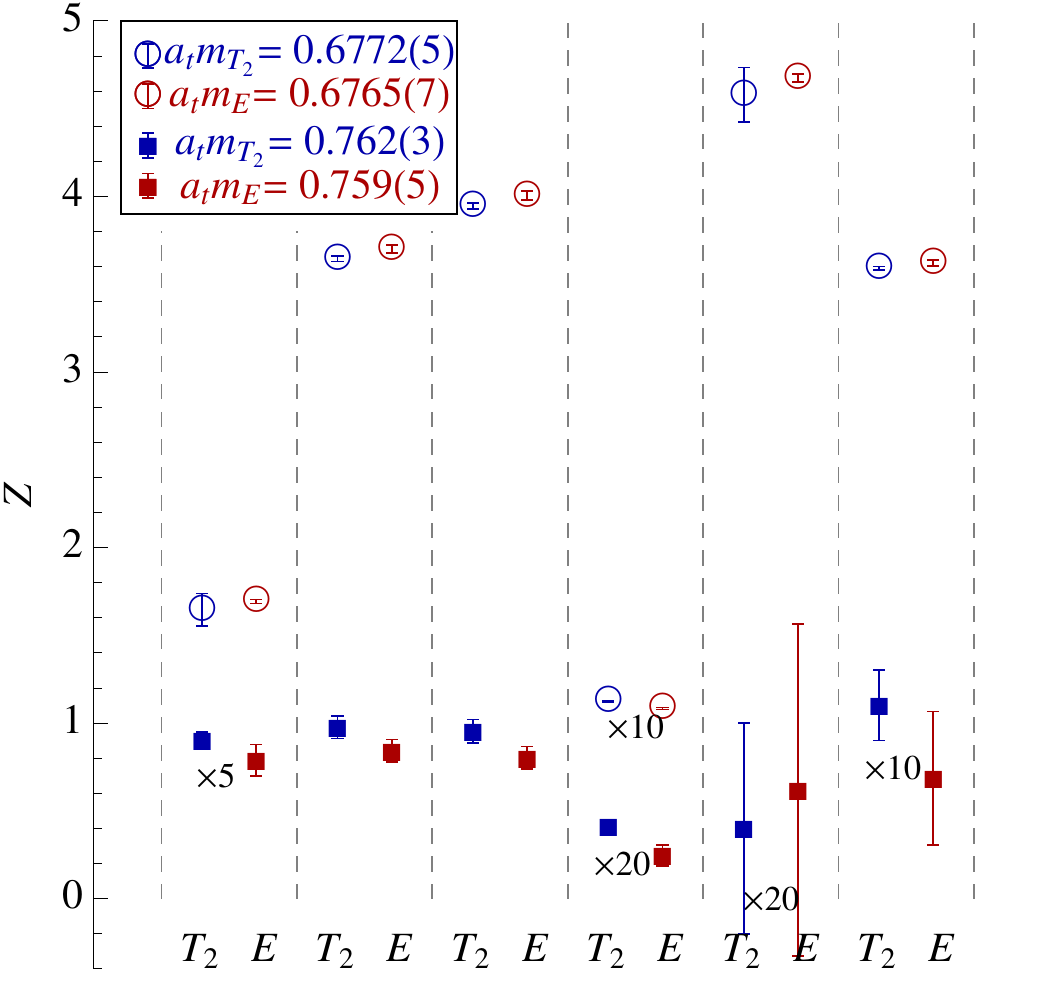}\quad\quad &
\includegraphics[height=0.36\textwidth]{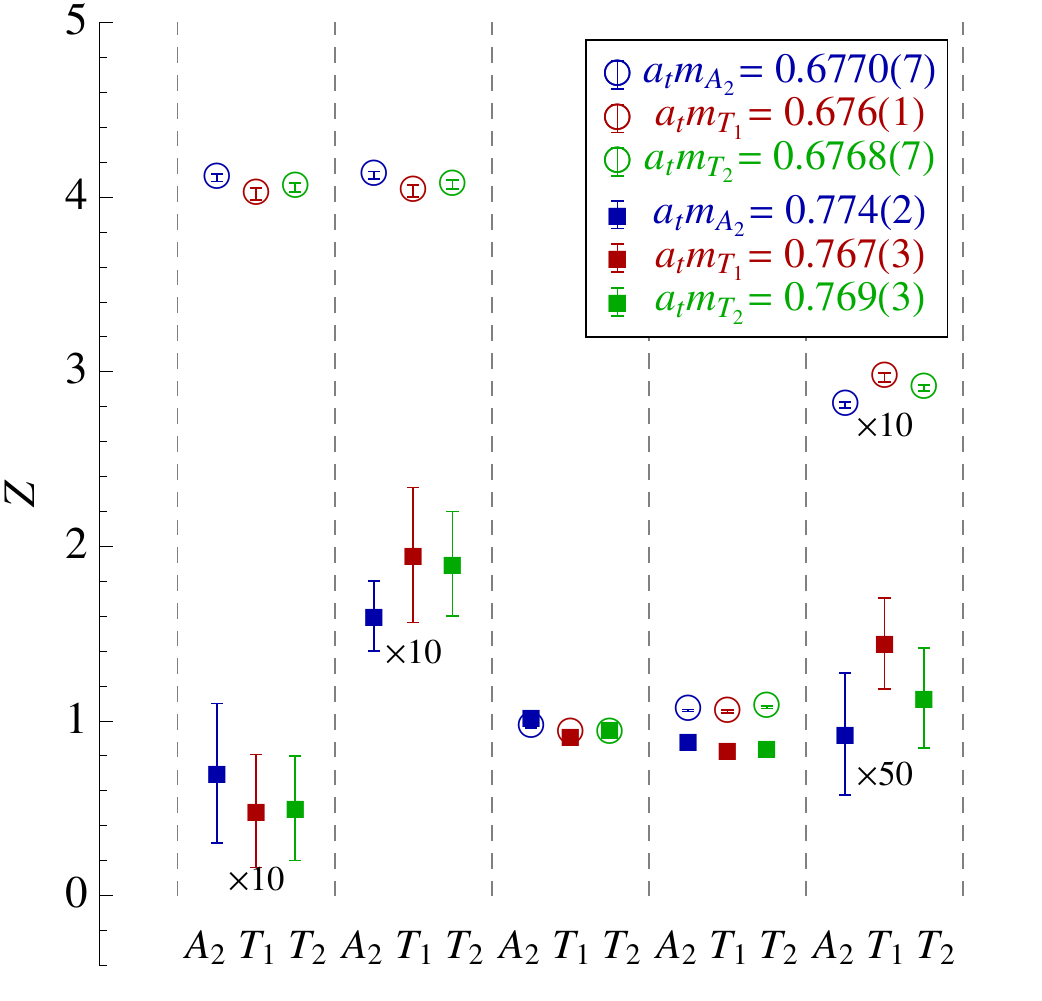} &
\includegraphics[height=0.36\textwidth]{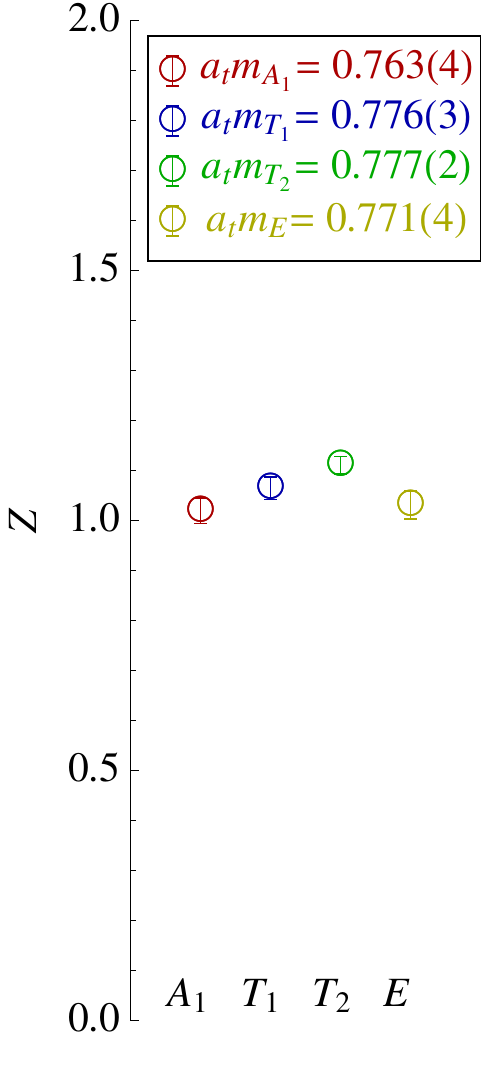}\\
$J=2$ & $J=3$ & $J=4$
\end{tabular}
\caption{A selection of $Z$-values for states conjectured to have $J=2, 3, 4$ in irreps 
$\Lambda^{--}$ on the $24^3$ volume.
From left to right, the operators in the left panel are 
$( a_1\times D^{[1]}_{J=1} )^{J=2}$, $( \rho_2 \times D^{[2]}_{J=2} )^{J=2}$, 
$( \rho \times D^{[2]}_{J=2} )^{J=2}$, $( a_0\times D^{[3]}_{J_{13}=2, J=2} )^{J=2}$, 
$( a_1\times D^{[3]}_{J_{13}=0, J=1} )^{J=2}$, $( a_1\times D^{[3]}_{J_{13}=2, J=1} )^{J=2}$; 
the operators in the middle panel are $(\rho_2 \times D^{[2]}_{J=2} )^{J=3}$, $(\rho \times D^{[2]}_{J=2} )^{J=3}$, $( a_0\times D^{[3]}_{J_{13}=2, J=3} )^{J=3}$, $( a_1\times D^{[3]}_{J_{13}=2, J=3} )^{J=3}$, $( b_1\times D^{[3]}_{J_{13}=1, J=2} )^{J=3}$; the operator in the right panel is $( a_1\times D^{[3]}_{J_{13}=2, J=3} )^{J=4}$.}
\label{Fig:Jmm_Z}
\end{center}
\end{figure*}

In Fig.~\ref{Fig:Jmm_Z} we show the $Z$-values for states conjectured to have 
$J=2$, 3 and 4 in the irreps $\Lambda^{--}$.  
It can be seen that there is generally good agreement between the 
$Z$-values extracted in different irreps although there are some deviations 
from exact equality.  The various discretisation effects have been discussed previously~\cite{Dudek:2010wm}.

\begin{figure*}[t]
\begin{center}
\includegraphics[width=1.0\textwidth]{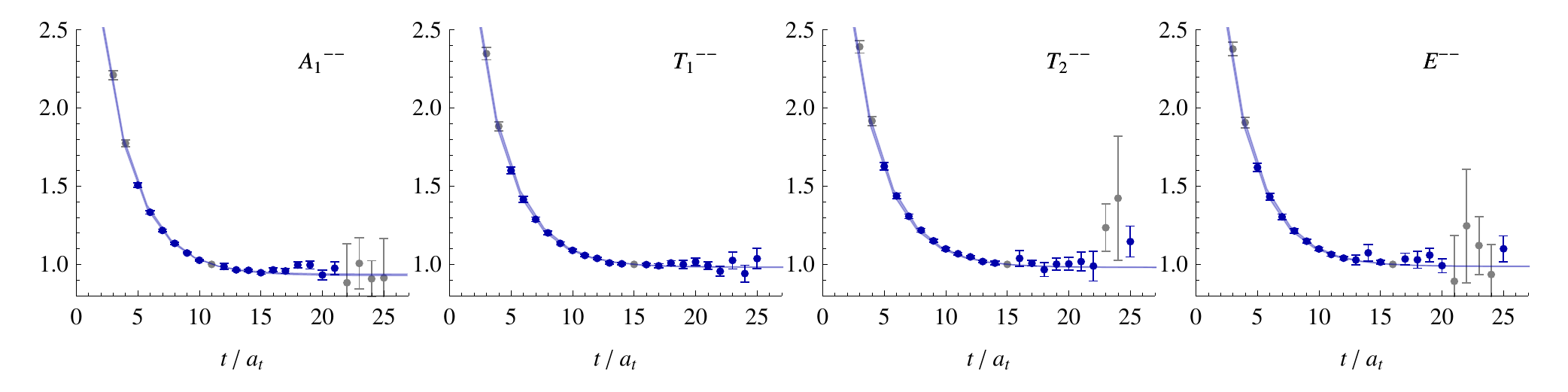}
\caption{A simultaneous fit to the four principal correlators of the lightest $4^{--}$ state on the $24^3$ volume using a common mass, $m_\mathfrak{n}$.  Plotted are $\lambda^{\mathfrak{n}}(t) \cdot e^{m_{\mathfrak{n}}(t-t_0)}$ data and the fit; the grey points are not included in the fit.}
\label{Fig:JointFit_4mm} 
\end{center}
\end{figure*}

For the $J \ge 2$ states which have components distributed across different irreps, the question then arises as to what we should quote as our estimate of the mass of the state.  The mass obtained from fits to principal correlators in each irrep can be different due to discretisation effects and fitting fluctuations.  To minimize the fluctuations caused by changes in the fitting of the principal correlators we perform a joint fit to the principal correlators~\cite{Dudek:2010wm}.  There is a common mass, $m_\mathfrak{n}$, in the fit but we allow the second exponential in the principal correlator [Eq.~\eqref{equ:fitprincorr}] in each irrep to be different and so the fit parameters are $m_\mathfrak{n}$, $\{m_\mathfrak{n}'^{\Lambda}\}$ and $\{A_\mathfrak{n}^{\Lambda}\}$.  Our final results are determined from such joint fits and we find they are generally reasonable, with $\chi^2/N_{\text{dof}} \sim 1$.  In Fig.~\ref{Fig:JointFit_4mm}, as an example, we present the joint fit of the lightest $4^{--}$ state in the four irreps $A_1^{--}$, $T_1^{--}$, $T_2^{--}$ and $E^{--}$.

In summary, we have demonstrated that the $Z$-values of our carefully-constructed subduced operators can be used to identify the continuum spin of the extracted states.  This was shown in the light meson sector~\cite{Dudek:2010wm} and here we have tested the method in an explicit calculation of the charmonium spectrum.  When we present our final spin-assigned spectra, for $J \ge 2$ states we use the results of joint fits to the principal correlators.

\section{Stability of the extracted spectrum}
\label{sec:systematics}

In this section we present the results of some systematic tests of the stability of the extracted spectrum.  We consider variations of the reference time-slice, $t_0$, used in the variational method, the number of distillation vectors and changes to the spatial clover coefficient, $c_s$.

\subsection{Variational analysis and $t_0$}
\label{subsec:t0}

In our implementation of the variational method the choice of a suitable $t_0$ is guided by how well the correlator is reconstructed, as described in Ref.~\cite{Dudek:2010wm}.  In this study we find that typically $t_0/a_t \sim 12 - 15$ is appropriate, depending on the irrep being considered.  

\begin{figure}[t]
\begin{center}
\includegraphics[width=1.0\textwidth]{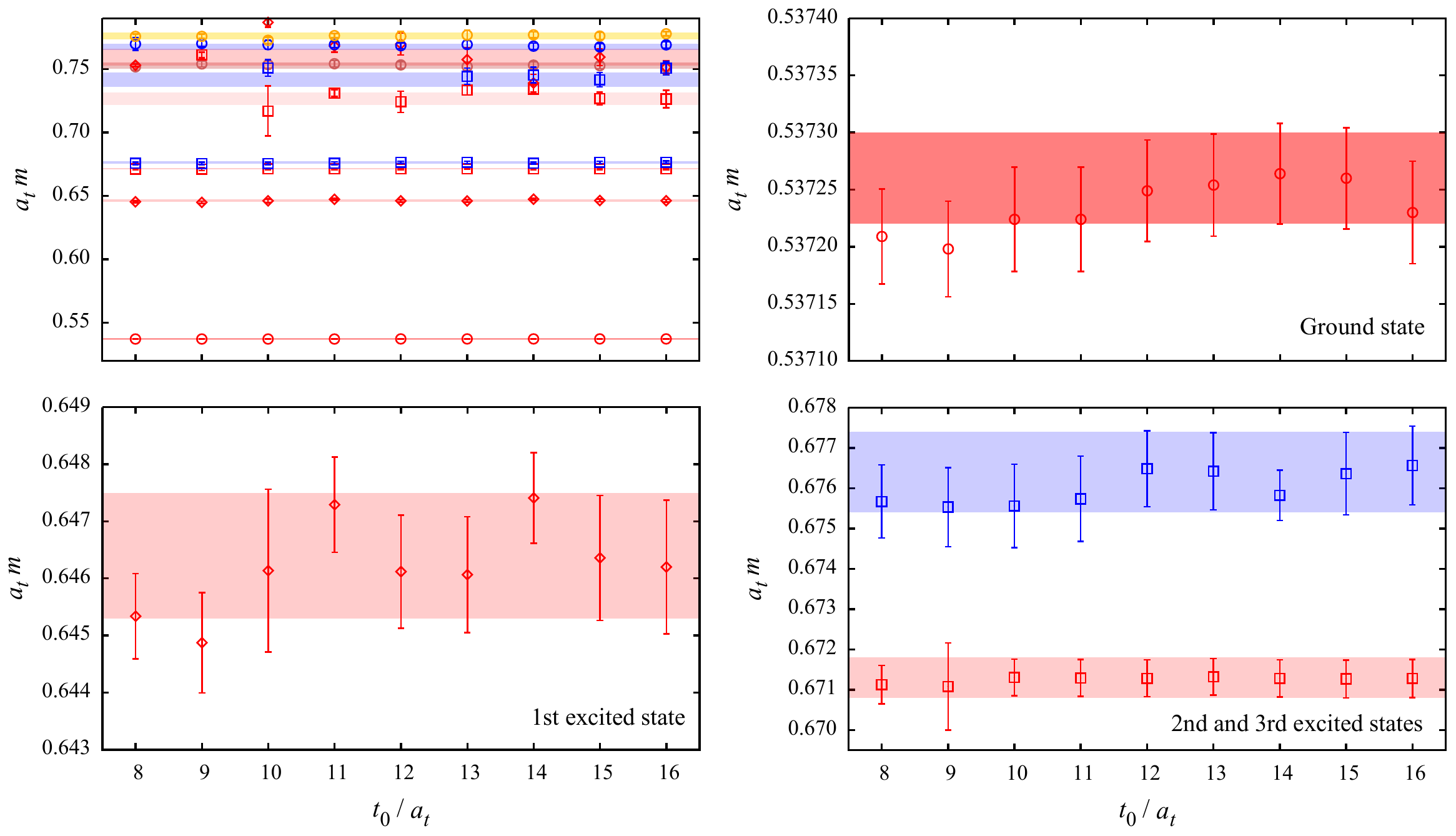}
\caption{Extracted mass spectrum as a function of $t_0$ for the lower-lying states in the $T_1^{--}$ irrep on the $24^3$ volume; the top left pane shows the lightest ten states and the other panes show the lightest four states in more detail.  Horizontal bands show the masses extracted at $t_0 = 15$ with the width $\pm1\sigma$ from the mean.  For large enough but not too large $t_0$ the spectrum is seen to be stable under changes in $t_0$. }
\label{fig:masst0}
\end{center}
\end{figure}

Fig.~\ref{fig:masst0} shows the variation with $t_0$ of the lightest ten extracted masses in the $T_1^{--}$ irrep on the $24^3$ ensemble.  It can be seen that for large enough but not too large $t_0$, the extracted spectrum is stable under variations in $t_0$.  We suggest that this stability is in a large part due to the inclusion of a second exponential term in Eq.~\eqref{equ:fitprincorr}.  This term can `mop up' other states which leak into the principal correlator because our operator basis is finite.  We find that typically the mass in this second exponential, $m'$, is larger than that of all of the extracted states.  As $t_0$ increases, the contribution of this second exponential falls rapidly due to it having a smaller amplitude, $A$, and a larger $m'$.

\begin{figure}[t]
\begin{center}
\includegraphics[width=0.7\textwidth]{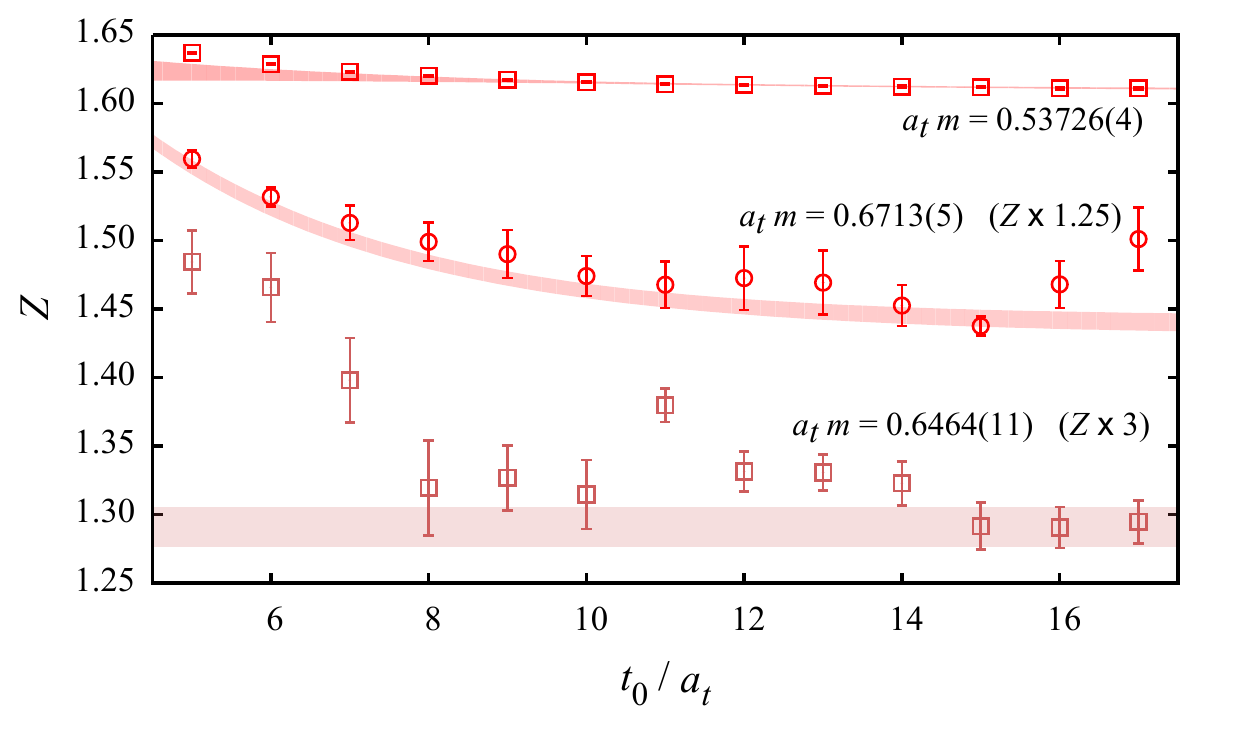}
\caption{Extracted overlaps, $Z$, with operator $(a_1 \times D^{[1]}_{J=1})^{J=1}_{T_1}$ as a function of $t_0$ for the lightest three $J=1$ states in the $T_1^{--}$ irrep on the $24^3$ volume.  The $Z$-values for the 1st and 2nd excited states have been arbitrarily scaled by factors of 3 and 1.25 respectively to fit on the plot.  Coloured bands show fits to an exponential plus a constant or a constant over an appropriate range of $t_0$.  The $Z$-values are seen to plateau for sufficiently large $t_0$ but can show significant curvature at small $t_0$.}
\label{fig:Zt0_spin1}
\end{center}
\end{figure}

\begin{figure}[t]
\begin{center}
\includegraphics[width=0.7\textwidth]{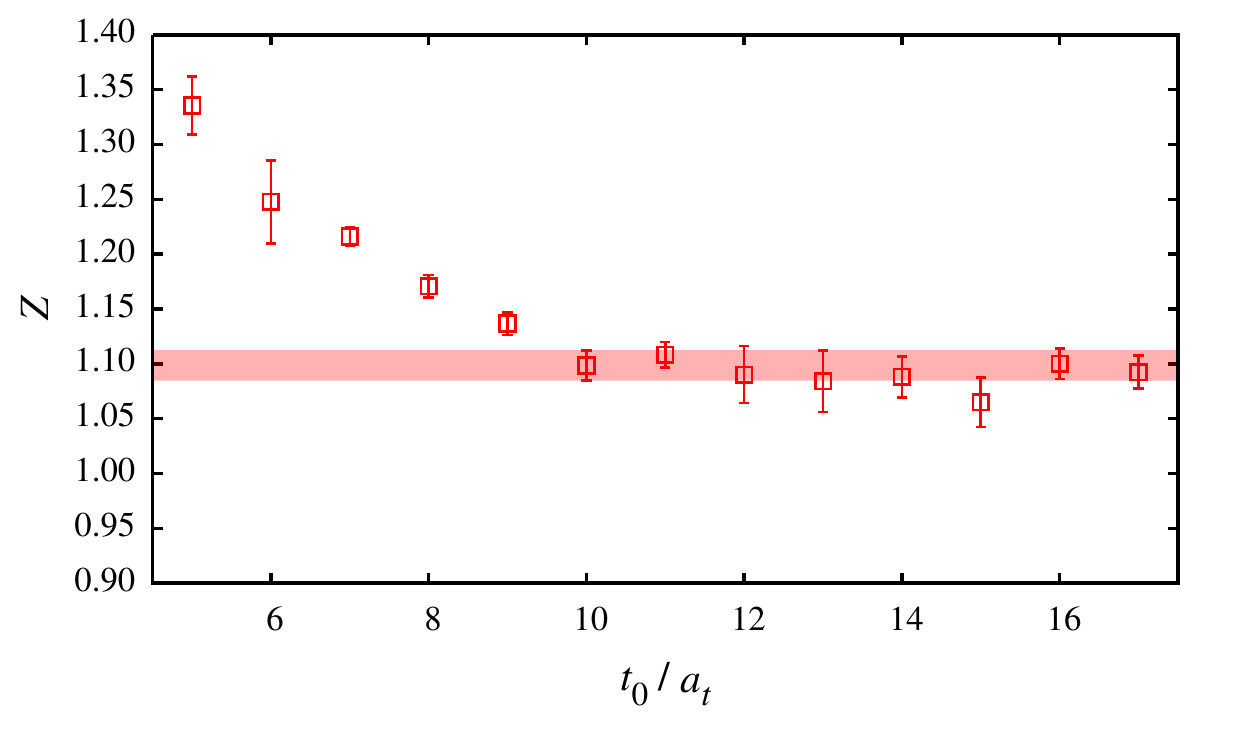}
\caption{Extracted overlaps, $Z$, with operator $(a_1 \times D^{[3]}_{J_{13}=2,J=3})^{J=4}_{T_1}$ as a function of $t_0$ for the lightest $J=4$ state in the $T_1^{--}$ irrep on the $24^3$ volume.  The horizontal band shows a fit to a constant over an appropriate range of $t_0$.  The $Z$-value is seen to plateau for large $t_0$ but shows significant curvature at small $t_0$.}
\label{fig:Zt0_spin4}
\end{center}
\end{figure}

In Figs.~\ref{fig:Zt0_spin1} and \ref{fig:Zt0_spin4} we show the extracted overlaps as a function of $t_0$ for a selection of the states in the $T_1^{--}$ irrep.  We observe that the overlaps show somewhat more sensitivity to $t_0$ than the masses.  This might be expected because the $Z$-values are related to the eigenvectors of the generalised eigenvalue problem; these are likely to be strongly effected by a `wrong' orthogonality when the number of contributing states is larger than the number of operators in the basis.

In summary, we conclude that our implementation of the variational method gives a reliable extraction of the spectrum if $t_0$ is chosen appropriately.  The masses extracted show very little variation with $t_0$ whereas the $Z$-values can show a more significant dependence on $t_0$.  This was also observed in a study of the spectrum of light isovector mesons~\cite{Dudek:2010wm}.

\subsection{Varying the number of distillation vectors}
\label{subsec:nvecs}

As described in Section \ref{sec:methods}, in the distillation method there is a choice of how many eigenvectors of the Laplacian, $\Nvecs$, to include in Eq.~\eqref{equ:distillation}.  Using fewer distillation vectors reduces the computational cost but using too few results in over smearing and limits our ability to extract high-spin and excited states~\cite{Dudek:2010wm}; the unsmeared limit corresponds to $\Nvecs = 3(L/a_s)^3$ where 3 corresponds to the number of colours.  To get the same smearing operator on a larger volume requires $\Nvecs$ to be scaled by a factor equal to the ratio of the spatial volumes~\cite{Peardon:2009gh}.  Therefore, to optimise the efficiency of these calculations, it is important to use the smallest number of distillation vectors for which the states of interest can be extracted reliably.

\begin{figure}[t]
\begin{center}
\includegraphics[width=0.7\textwidth]{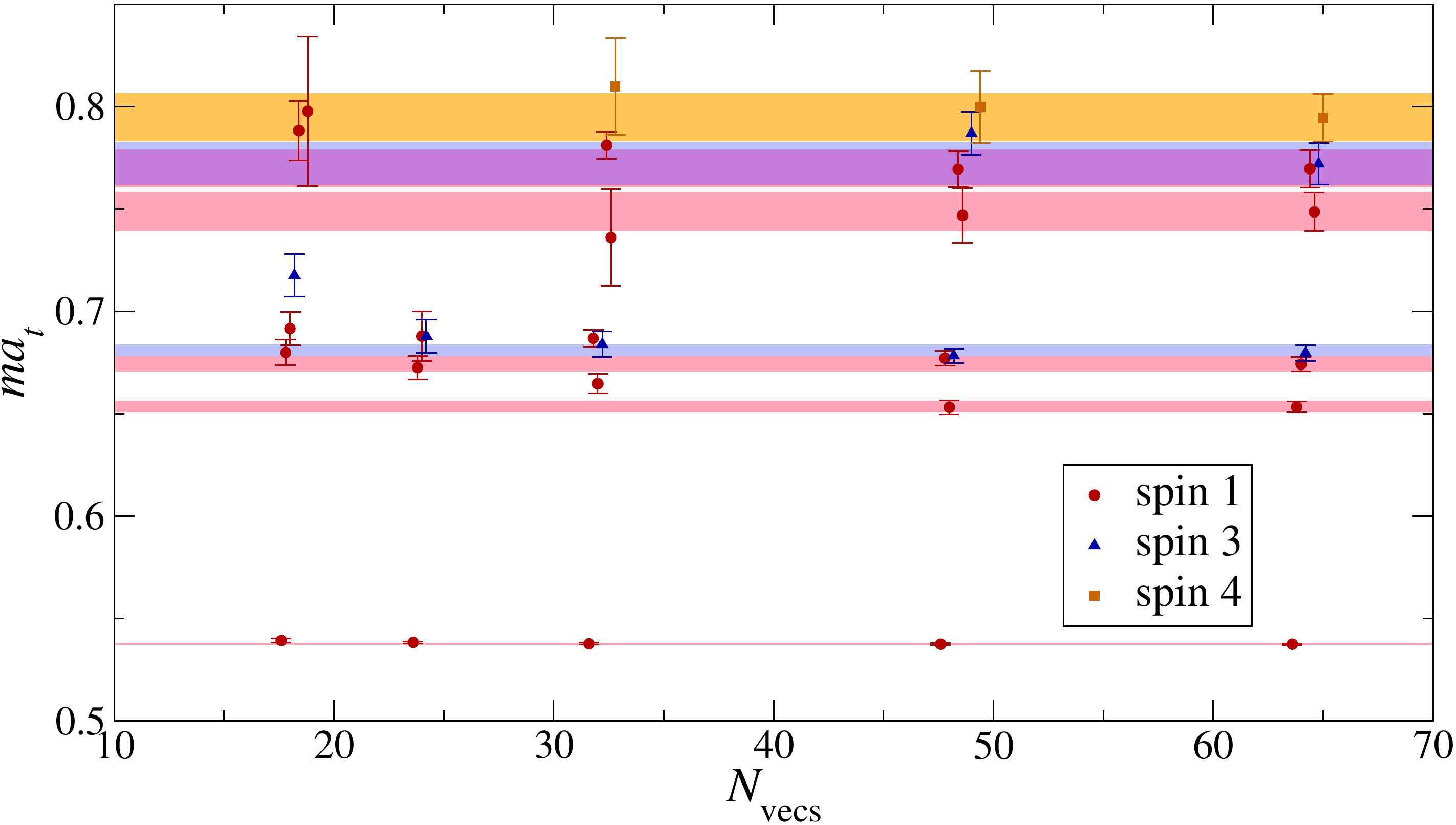}
\caption{Extracted spectrum for the $T_1^{--}$ irrep on the $16^3$ volume as a function of the number of distillation vectors, $\Nvecs$, using lower statistics than the main $16^3$ results.  Horizontal bands correspond to the masses extracted with $\Nvecs=64$ and give the 1 sigma range on either side of the mean.  The spectrum is observed to be stable for $\Nvecs \gtrsim 48$.}
\label{fig:neig}
\end{center}
\end{figure}

In Fig.~\ref{fig:neig} we show the low-lying states in the $T_1^{--}$ irrep on the $16^3$ volume as a function of the number of distillation vectors used.  These correlators were calculated using lower statistics, 90 gauge-field configurations and 6 time-sources per configuration, compared to the main $16^3$ results and so the extracted spectrum at $\Nvecs=64$ is not identical to that in the full results presented later.  The spin of some states for $\Nvecs \le 32$ could not be reliably identified and these are not included in the figure.   The spectrum is reasonably stable for $\Nvecs \gtrsim 48$, although the statistical precision reduces slightly with $\Nvecs = 48$ compared to $\Nvecs = 64$.  The quality of the spectrum degrades for fewer vectors and this is particularly apparent for the excited and high-spin states.  A possible explanation for this behaviour was given in Ref.~\cite{Dudek:2010wm}.

In summary, the number of distillation vectors must be sufficiently large in order to reliably extract high-spin and excited states, and once the number of vectors is large enough the spectrum is stable under changes in the number.  These results are consistent with those of Ref.~\cite{Dudek:2010wm}.  The observation that $48$ vectors are sufficient on the $16^3$ volume suggests that $48 \left(\tfrac{24}{16}\right)^3 = 162$ vectors is reasonable for the $24^3$ volume.

\subsection{${\cal O}(a_s)$ effects and the hyperfine splitting}
\label{subsec:c_sw}

An accurate determination of the hyperfine splitting between the ground-state 
pseudoscalar and vector mesons made from two heavy quarks 
has been a long-standing problem for lattice investigations. The determination 
of this energy is known to be delicately sensitive to artefacts arising from 
the discretisation of the Dirac operator on the lattice. The result for the 
$J/\psi-\eta_c$ splitting determined on our $16^3$ ensemble using tree-level 
${\cal O}(a)$ improvement is 80(2) MeV, consistent with that obtained 
on the $24^3$ ensemble and substantially undershooting the physical value of 117 MeV. 
This determination neglects
the effects of the disconnected Wick graphs contributing to the correlation 
function of charmonium mesons. These diagrams are not expected to change the 
hyperfine splitting substantially \cite{Levkova:2010ft} due to OZI suppression 
and their absence is 
unlikely to explain in full the discrepancy noted above. We have made some first 
tests of the effect of disconnected diagrams in our calculation and come to 
the same preliminary conclusion. One other source of systematic uncertainty is 
the use of up and down quarks heavier than their physical values in our 
lattice action. Other lattice calculations \cite{Bali:2011dc} suggest 
this will also result in only a small effect which will not explain the discrepancy 
we see. 

\begin{figure}[t]
\begin{center}
  \includegraphics[width=0.7\textwidth]{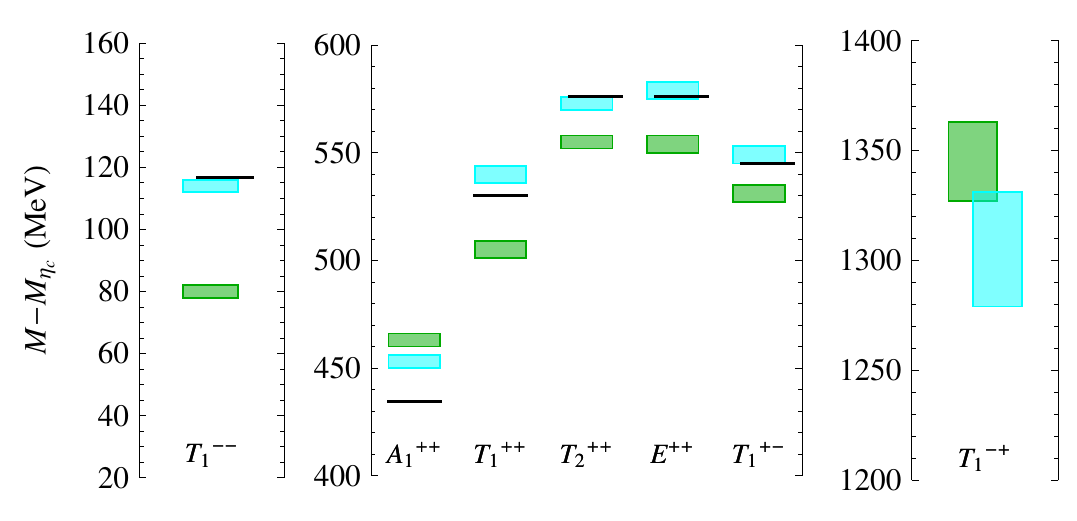}
  \caption{The mass shifts, measured on the $16^3$ ensemble, for the lowest-lying states in a selected set of charmonium lattice irreps as 
$c_s$, the size of the chromomagnetic clover term, is varied from tadpole-improved tree-level value of $c_s=1.35$ (green) to $c_s=2$ (cyan). Masses presented here are measured relative to $M_{\eta_c}$. 
Experimental data is shown as solid lines; note that experimentally the $\chi_{c0}$ has a significant hadronic width.}
  \label{fig:tuned_cs_spectrum}
\end{center}
\end{figure}

The lattice action for charm quark fields used in this study is 
${\cal O}(a)$-improved at tree level in tadpole-improved perturbation theory. 
It would be expected that a non-perturbative determination 
\cite{Luscher:1996ug} of action parameters would yield a larger
value of $c_s$, the coefficient of the spatial Pauli-like term, 
$\bar\psi \sigma_{ij} F_{ij} \psi$, appearing in the 
lattice action, than the value $c_s=1.35$ we use. Studies show that increasing the value of $c_s$ yields 
a larger hyperfine splitting at finite lattice spacing. To investigate the 
effect of a larger value of $c_s$, a study of some charmonium states using 
$c_s=2$ was carried out on the $16^3$ volume. To be clear, the choice of $c_s=2$ is an ad hoc one, 
chosen from the consideration that a non-perturbative determination is expected to be larger 
than the tadpole-improved tree-level value which in our simulations is $c_s=1.35$. 

The lowest lying $0^{-+}$, $1^{--}$, 
$(0,1,2)^{++}$ and $1^{-+}$ states were computed, corresponding to the 
lightest S and P-wave states in a quark model and the lightest exotic, and 
Fig.~\ref{fig:tuned_cs_spectrum} shows these masses with the $\eta_c$-mass subtracted.
Choosing $c_s=2$ gives a mass difference between the $J/\psi$ and $\eta_c$ of 114(2) MeV, very
close to the experimental value and also 
leads to significantly better agreement in the P-wave system. 
The spin-two $\chi_{c2}$ state is split on the lattice across two irreps, 
$T_2^{++}$ and $E^{++}$. For both values of $c_s$, these two states are 
degenerate within the statistical precision of this study. This is still consistent
with the explanation that the hyperfine splitting is underestimated due to ${\cal O}(a)$ lattice
artefacts. Since all dimension-five operators consistent with lattice 
symmetries do not break the continuum rotation group, any differences between
these lattice states should only arise at ${\cal O}(a^2)$ in a Symanzik 
expansion and so would be expected to be small. 

The lightest charmonium exotic, the $1^{-+}$ state, shows mild dependence on 
$c_s$. This suggests the energies of such states in our study should be close 
to their continuum values. 
Note that the results presented later in Section~\ref{sec:results} use the tree-level 
tadpole-improved value of $c_s$, not the boosted version of this test.  
This computation does however enable us to assign an approximate scale of 40 
MeV to the size of the systematic uncertainty arising from the leading-order 
${\cal O}(a_s)$ correction and observe that it does not increase for the higher-lying states in the spectrum where this has been tested.  However, a more detailed investigation of these discretisation effects has not yet been performed and would be required for more complete control of this systematic uncertainty.

\section{Results}
\label{sec:results}

In this section we present the results from our lattice calculations.  We first give the results by lattice irrep and compare the spectra obtained on the two lattice volumes.  We then show the final results with states labelled by their continuum quantum numbers, $J^{PC}$.  Following on from this, in Section \ref{sec:discussion} we interpret the results and discuss the hybrid phenomenology suggested by them.

\begin{figure}[tb]
\begin{center}
\includegraphics[width=0.9\textwidth]{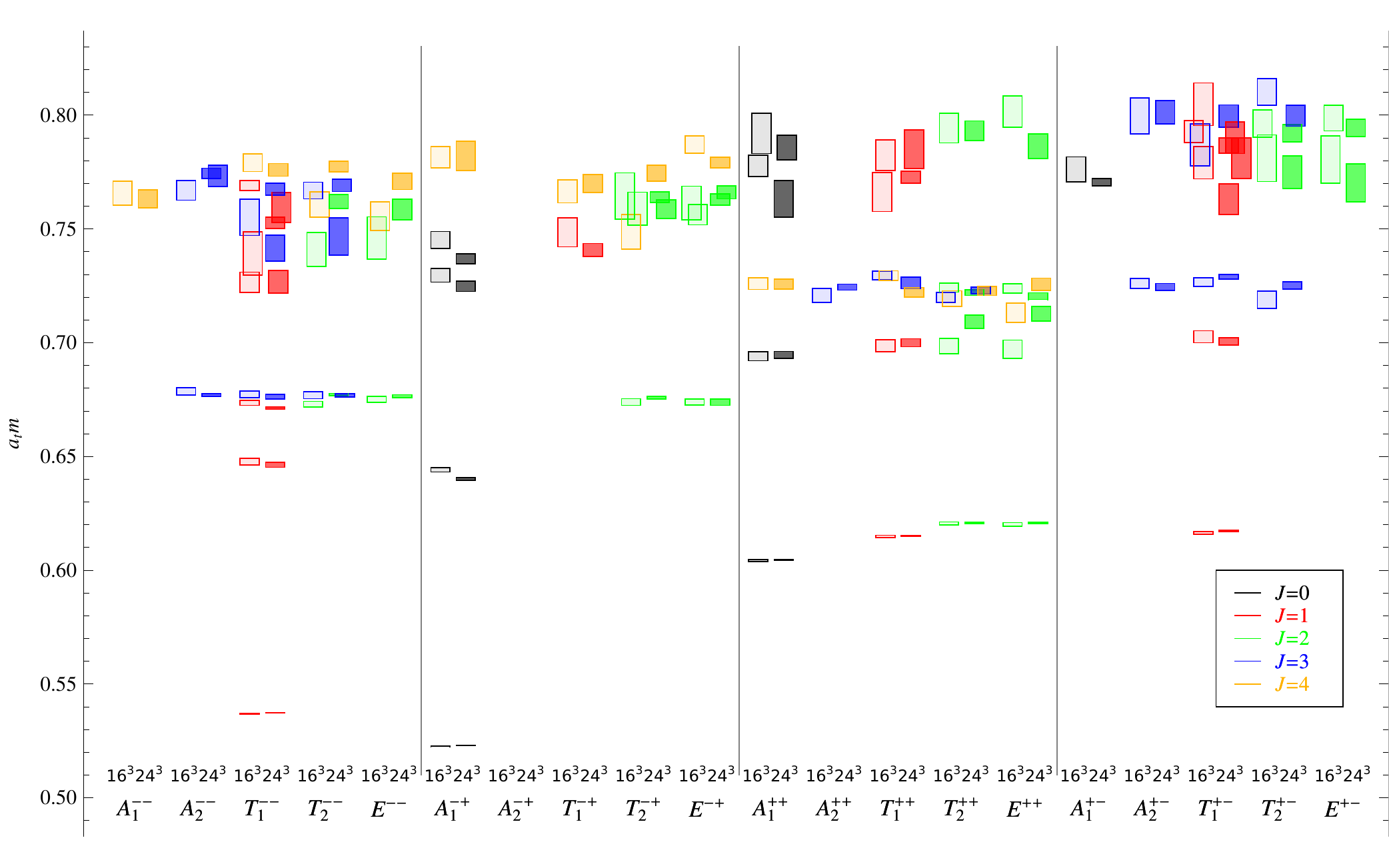}
\caption{Extracted spectra by lattice irrep, $\Lambda^{PC}$, on the $16^3$ (lighter shading) and $24^3$ (darker shading) volumes.  The vertical size of each box gives the one sigma statistical uncertainty on either side of the mean and the colour coding indicates the continuum spin.}
\label{fig:latticeirreps}
\end{center}
\end{figure}

\subsection{Results by lattice irrep and volume comparison}
\label{sec:results:latticeirreps}

In Fig.~\ref{fig:latticeirreps} we show the results of the variational analysis for each lattice irrep, $\Lambda^{PC}$, with the spectra obtained from the two volumes side by side.  The one sigma statistical uncertainty on either side of the mean, as determined from the variational analysis, is indicated by the vertical size of the boxes.  The colour coding gives the continuum spin determined by considering the operator overlaps as described in Section~\ref{sec:spin}.  States identified as $J=0$ are coloured black, $J=1$ are red, $J=2$ are green, $J=3$ are blue and $J=4$ are gold.  

\begin{figure}[t]
\begin{center}
\includegraphics[width=0.9\textwidth]{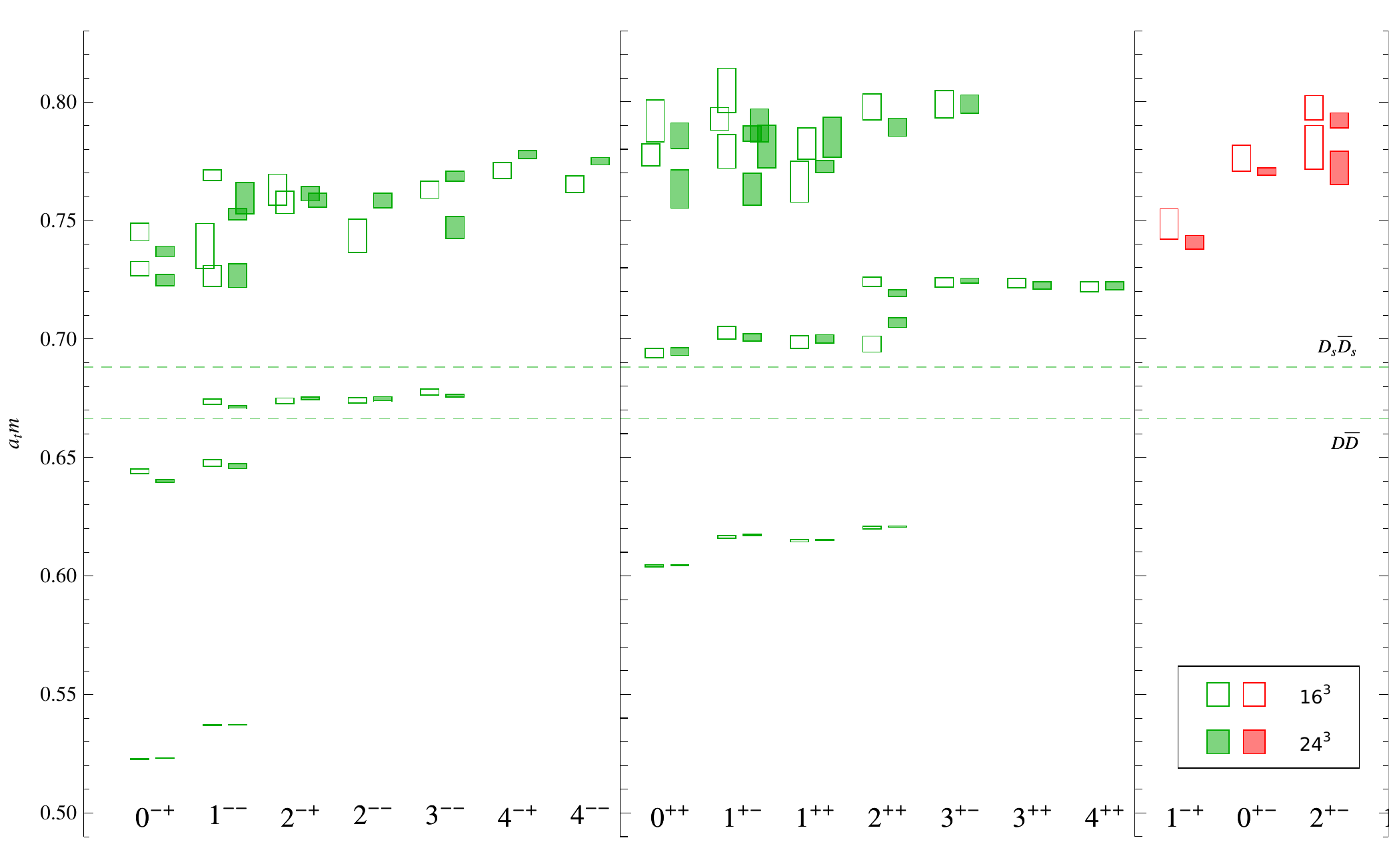}
\caption{The extracted spin-identified charmonium spectrum labelled by $J^{PC}$; the $16^3$ (open boxes) and $24^3$ (filled boxes) volumes agree well.  The vertical size of the boxes represents the one sigma statistical uncertainty on either side of the mean.  The dashed lines indicate the lowest non-interacting $D\bar{D}$ and $D_s\bar{D}_s$ levels using the $D$ and $D_s$ masses measured on the $16^3$ ensemble.}
\label{Fig:Summary_VolCompare}
\end{center}
\end{figure}

By comparing the overlaps between different irreps, we are able to match levels which correspond to the different components of a state with definite continuum spin subduced into lattice irreps; generally we see no significant discrepancy between the masses determined in different irreps.  The dense spectrum of states at and above $a_t m \sim 0.67$ would appear to be impossible to disentangle using solely the extracted masses.  An advantage of using anisotropic lattices is that all masses we consider have $a_tm<1$, which is not easily achieved for charmonium on isotropic lattices.  The spatial lattice spacing is $a_s \sim 0.12$~fm and this is fine enough for an effective restoration of rotational symmetry at the hadronic scale, shown by the fact that we can successfully determine the continuum spin of extracted states.

In Fig.~\ref{Fig:Summary_VolCompare} we show the spin-identified spectrum labeled by continuum $J^{PC}$ with the results from the two volumes presented side by side; only the well-determined low-lying states are included.  The dashed lines indicate the lowest thresholds for open charm decay, the non-interacting $D\bar{D}$ and $D_s\bar{D_s}$ levels with the $D$ and $D_s$ masses measured on the $16^3$ ensemble.  We observe that there is no significant difference between the two volumes at the level of the statistical uncertainties, even above the open-charm thresholds, evidence that we are not seeing multi-hadron states in our extracted spectra.  We used higher statistics for the $24^3$ calculation compared to the $16^3$ calculation (see Section \ref{subsec:action}).  On the $24^3$ volume we can reliably extract and identify the spin of more states and determine their masses with higher statistical precision, and we will therefore quote these as our final results.

In summary, we are able to interpret our spectra in terms of single mesons of definite continuum spin subduced into various lattice irreps.  This, along with the lack of any significant volume dependence, is evidence that we are not observing multi-hadron states in our extracted spectra; the distribution of such states across irreps would have a different pattern.  Further evidence is provided by observing that we do not extract energy levels where we would expect non-interacting two-meson levels to appear.  These points have been discussed in detail in Ref.~\cite{Dudek:2010wm}.

\subsection{Final spin-identified spectrum}
\label{sec:results:final}

\begin{figure}[t]
\begin{center}
\includegraphics[width=1.0\textwidth]{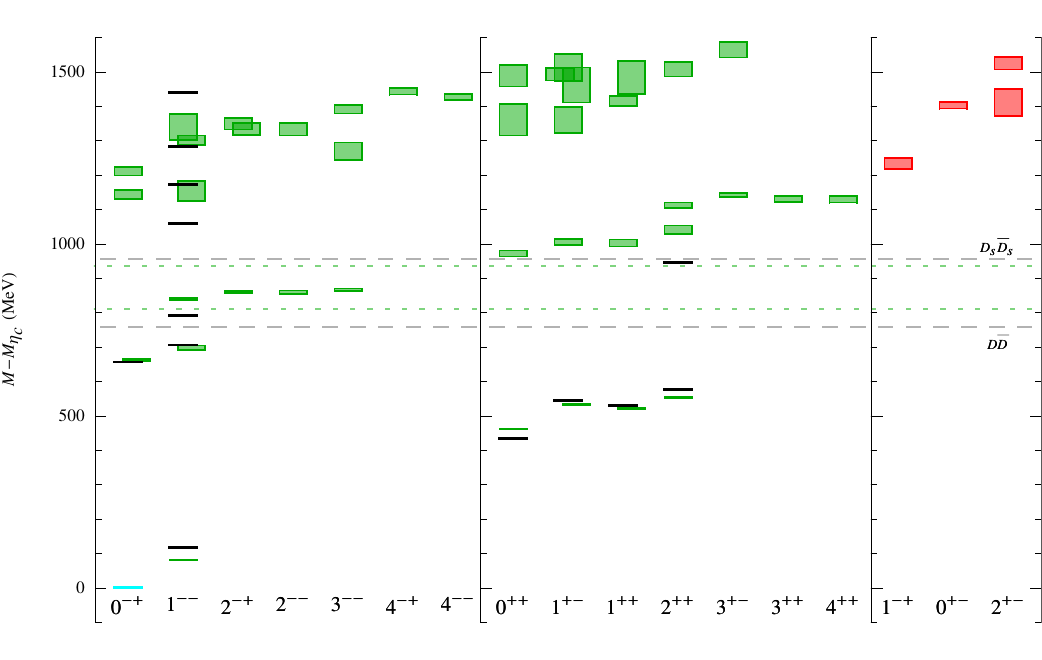}
\caption{Summary of the charmonium spectrum up to around $4.5~\text{GeV}$ labelled by $J^{PC}$.  The red and green boxes are the masses calculated on the $24^3$ volume; black lines are experimental values from the PDG~\cite{PDG2011}.  We show the calculated (experimental) masses with the calculated (experimental) $\eta_c$ mass subtracted.  The vertical size of the boxes represents the one sigma statistical uncertainty on either side of the mean.  The dashed lines indicate the lowest non-interacting $D\bar{D}$ and $D_s\bar{D}_s$ levels using the $D$ and $D_s$ masses measured on the $16^3$ ensemble (fine green dashing) and using the experimental masses (coarse grey dashing).}
\label{Fig:Summary} 
\end{center}
\end{figure}

Our final results, the well-determined states on the $24^3$ volume, are shown in Fig.~\ref{Fig:Summary} along with experimental masses taken from PDG summary tables~\cite{PDG2011}; these results are also tabulated in Appendix~\ref{app:resultstable}.  The $X(3872)$ is not shown because its $J^{PC}$ ($1^{++}$ or $2^{-+}$) has not been determined experimentally.  In the plot, the calculated (experimental) mass of the $\eta_c$ has been subtracted from the calculated (experimental) masses in order to reduce the systematic error from the tuning of the bare charm quark mass (see Section \ref{sec:lattices}).  The dashed lines indicate the lowest non-interacting $D\bar{D}$ and $D_s\bar{D_s}$ levels using the $D$ and $D_s$ masses measured on the $16^3$ ensemble (fine green dashing) and using the experimental masses (coarse grey dashing).  We note that higher up in the spectrum the states become less well determined.  This is in part because, although we have a relatively large number of operators in each channel, the basis is still restricted in size.  In order to better determine these states we would need to include operators with different radial structures and, in order to determine states with $J \ge 5$, different angular structures, for example higher numbers of derivatives.

In the following section we interpret our results and compare with the experimental situation, but before that we discuss some other lattice calculations.  Ref.~\cite{Dudek:2007wv} presented a pioneering quenched lattice QCD calculation of the excited charmonium spectrum which was interpreted further in Ref.~\cite{Dudek:2008sz}.  In comparison, apart from being a dynamical calculation, the current work uses a larger basis of operators and these operators have been constructed so as to facilitate the identification of the continuum spin of the extracted states.  This means that we can precisely extract a much larger number of excited states and states with high spin, and reliably identify their continuum $J^{PC}$.

Results from an $N_f = 2$ calculation (dynamical up and down quarks but quenched strange quarks) of excited charmonia are presented in Ref.~\cite{Bali:2011rd} using isotropic lattices with a few different lattice spacings and $m_{\pi}$. They also consider mixing with light mesons in the pseudoscalar channel and mixing with some low-lying multi-mesons states in a few channels. In addition, preliminary results from dynamical ($N_f = 2+1$) calculations of excited charmonium spectra have been presented in Ref.~\cite{Bali:2011dc}. A range of pion masses and lattice spacings are considered and chiral and continuum extrapolations are attempted. However, both these references use a smaller operator basis compared to ours making a robust identification of the continuum spin of the extracted states difficult. Therefore a limited number of states can be reliably extracted and in particular the lightest exotic hybrid ($1^{-+}$) is difficult to unambiguously disentangled from a non-exotic spin four state.

\section{Interpretation of the results}
\label{sec:discussion}

In this section we give some interpretation of our results, compare with the experimental situation and then discuss hybrid meson phenomenology in more detail.  Many of the states with non-exotic $J^{PC}$ in Fig.~\ref{Fig:Summary} appear to follow the $n^{2S+1}L_J$ pattern predicted by quark potential models, where $S$ is the spin of the quark-antiquark pair, $L$ is the relative orbital angular momentum, $J$ is the total spin of the meson and $n$ is the radial quantum number.  The quantum numbers of the members of various quark-antiquark \emph{supermultiplets} are listed in Table \ref{table:multiplets}.  The $^{2S+1}L_J$ assignments for our extracted states are determined by considering the operator-state overlaps~\cite{Dudek:2008sz}.  

\begin{table}[t]
\begin{center}
\begin{tabular}{lccc}
$J^{P_g C_g}_g$ & $L$ & $S$ & $J^{PC}$  \\
\hline
\hline
\multirow{2}{*}{--} &
\multirow{2}{*}{$0(S)$}
   & 0 & $0^{-+}$ \\
 & & 1 & $1^{--}$ \\
\hline
\multirow{2}{*}{--} &
\multirow{2}{*}{$1(P)$}
   & 0 & $1^{+-}$ \\
 & & 1 & $(0,1,2)^{++}$ \\
\hline
\multirow{2}{*}{--} &
\multirow{2}{*}{$2(D)$}
   & 0 & $2^{-+}$ \\
 & & 1 & $(1,2,3)^{--}$ \\
\hline
\multirow{2}{*}{--} &
\multirow{2}{*}{$3(F)$}
   & 0 & $3^{+-}$ \\
 & & 1 & $(2,3,4)^{++}$ \\
\hline
\multirow{2}{*}{--} &
\multirow{2}{*}{$4(G)$}
   & 0 & $4^{-+}$ \\
 & & 1 & $(3,4,5)^{--}$ \\
\hline
\multirow{2}{*}{$1^{+-}$} &
\multirow{2}{*}{$0(S)$}
   & 0 & $1^{--}$ \\
 & & 1 & $(0,\mathbf{1},2)^{-+}$ \\
\hline
\multirow{2}{*}{$1^{+-}$} &
\multirow{2}{*}{$1(P)$}
   & 0 & $(0,1,2)^{++}$ \\
 & & 1 & $(\mathbf{0},1,1,1,\mathbf{2},\mathbf{2},3)^{+-}$ \\
\hline
\end{tabular}
\caption{Supermultiplets for quark-antiquark pairs with spin $S$ and relative orbital angular momentum $L$.  Also shown are some hybrid supermultiplets, discussed in Section~\ref{sec:discussion:hybrids}, where $J^{P_g C_g}_g$ are the quantum numbers of the gluonic excitation; exotic $J^{PC}$ are shown in bold.}
\label{table:multiplets}
\end{center}
\end{table}

In the left panel of Fig.~\ref{Fig:Summary} we show the negative parity states.  We observe the ground-state S-wave pair $[0^{-+}, 1^{--}]$ and at $M-M_{\eta_c} \sim$ 700~MeV the first excitation; a second excitation is observed at $M-M_{\eta_c} \sim$ 1150~MeV.  There is a complete D-wave set [$(1, 2, 3)^{--}$, $2^{-+}$] just above the $D\bar{D}$ threshold.  In the region above 1200~MeV there is a complete set of D-wave excitations and parts of a G-wave indicated by the presence of spin-4 states.  In addition, there are three states at $M-M_{\eta_c} \sim$ 1200 to 1300~MeV with $J^{PC} = (0, 2)^{-+}, 1^{--}$ which do not appear to fit with the pattern expected by quark models.  We note that these three states all have a relatively large overlap onto operators proportional to the gluonic field-strength tensor, something not observed for the states that fit into quark-antiquark supermultiplets.  For example, this $1^{--}$, the sixth state in the $T_1^{--}$ irrep in Fig.~\ref{Fig:Jmm_histo}, is the only state shown which has a large overlap onto the $(\pi \times D^{[2]}_{J=1})^{J=1}$ operator, an operator proportional to the chromomagnetic part of the field-strength tensor.  Following Ref.~\cite{Dudek:2010wm} we suggest that these `excess' states can be identified as non-exotic hybrid mesons and we return to this in Section~\ref{sec:discussion:hybrids}.

In the middle panel we show the positive parity states.  Below $D\bar{D}$ threshold there is a clear P-wave set [$(0, 1, 2)^{++}$, $1^{+-}$].  In the region of $M-M_{\eta_c} \sim$ 1000~MeV there is a P-wave excitation and, slightly higher, an F-wave [$(2, 3, 4)^{++}$, $3^{+-}$].  The band of states around $M-M_{\eta_c} \sim$ 1400~MeV probably contains part of the second excitation of the P-wave set and several non-exotic hybrids which lie above the lightest negative-parity hybrids; we will discuss these hybrid candidates in Section.~\ref{sec:discussion:hybrids}.

The states with exotic $J^{PC}$ are presented in the right panel.  These cannot consist solely of a quark-antiquark pair -- more degrees of freedom are necessary.  In general, there are several possible interpretations such as hybrid mesons where the gluonic field is excited, molecular mesons or tetraquarks.  The exotic states in our spectrum have significant overlap onto operators with non-trivial gluonic structure and this suggests that we can identify them as hybrid mesons.  Further, the lack of evidence for multi-hadron states in any channel, as discussed in Section~\ref{sec:results}, suggests that these exotics are not multi-hadron states.  The $1^{-+}$ is the lightest with $M-M_{\eta_c} \sim$ 1200~MeV and nearly degenerate with the three non-exotic hybrid candidates [$(0, 2)^{-+}$, $1^{--}$] observed in the negative parity sector.  Higher in the spectrum there is a pair of states, $(0, 2)^{+-}$, at $M-M_{\eta_c} \sim$ 1400~MeV and a second $2^{+-}$ state slightly higher again.  In Section \ref{sec:discussion:hybrids} we discuss the hybrid meson phenomenology suggested by our spectra.

\begin{figure}[t]
\begin{center}
\includegraphics[width=0.48\textwidth]{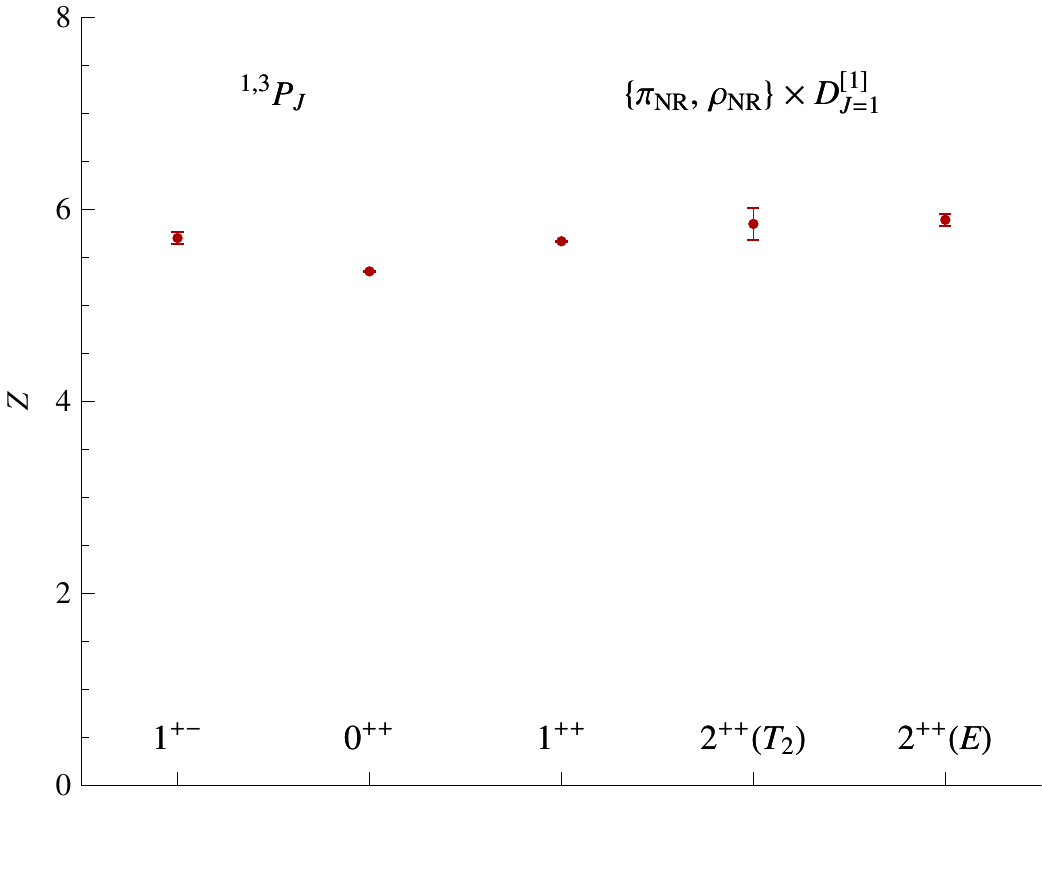} \hfill
\includegraphics[width=0.48\textwidth]{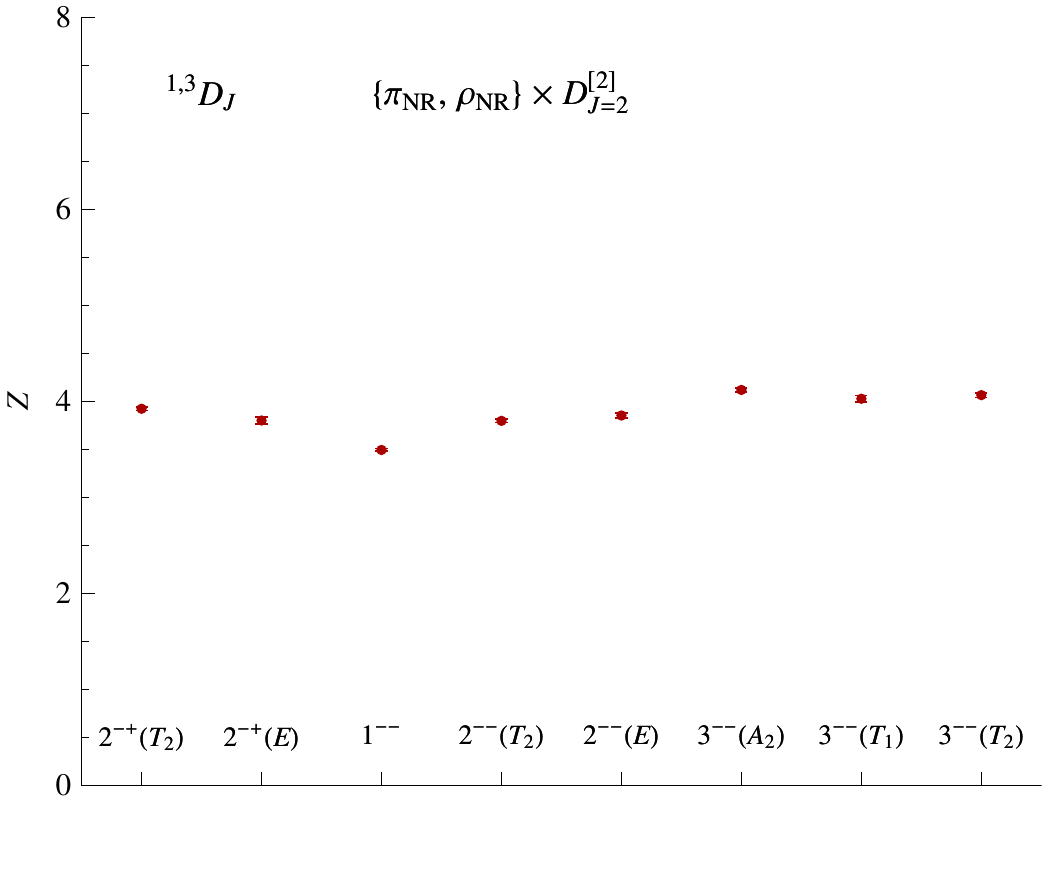}
\caption{Overlaps, $Z^\mathfrak{n}_i \equiv \langle \mathfrak{n} | \mathcal{O}^{\dagger}_i | 0 \rangle$, for the lightest P-wave (left panel) and D-wave (right panel) supermultiplets with operators $\{\pi,\rho\}_{\text{NR}}\times D_{J=1}^{[1]}$ and $\{\pi,\rho\}_{\text{NR}}\times D_{J=2}^{[2]}$ respectively.}
\label{fig:multiplets:conventional}
\end{center}
\end{figure}

In order to further test the assignment of states into the lightest P and D-wave supermultiplets, we follow Ref.~\cite{Dudek:2011bn} and compare operator-state overlaps for different states within a supermultiplet.  Consider the operators $(\pi_{\text{NR}} \times D_{J=1}^{[1]})^{J=1}$ with $J^{PC} = 1^{+-}$ and $(\rho_{\text{NR}} \times D_{J=1}^{[1]})^{J=0,1,2}$ with $J^{PC} = (0,1,2)^{++}$, where $\rho_{\text{NR}} = \frac{1}{2}\gamma_i (1 - \gamma_0)$ and $\pi_{\text{NR}} = \frac{1}{2}\gamma_5 (1 - \gamma_0)$.\footnote{these are for creation operators and we correct a typo appearing in the definition $\pi_{\text{NR}}$ in Ref.~\cite{Dudek:2011bn}}  These operators have the structure of a quark-antiquark pair in a gauge-covariant version of a P-wave with $S=0$ ($\pi_{\text{NR}}$) or $S=1$ ($\rho_{\text{NR}}$).  The operators $(\pi_{\text{NR}} \times D_{J=2}^{[2]})^{J=2}$ with $J^{PC} = 2^{-+}$ and $(\rho_{\text{NR}} \times D_{J=2}^{[2]})^{J=1,2,3}$ with $J^{PC} = (1,2,3)^{--}$ have the structure of a quark-antiquark pair in D-wave with $S=0$ or $S=1$.

In Fig.~\ref{fig:multiplets:conventional} we show the overlaps of these operators with states suggested to be members of the lightest P-wave (left panel) and D-wave (right panel) supermultiplets, and observe that these overlaps are significant.  More generally, we find that the overlap onto non-relativistic operator combinations~\cite{Dudek:2011bn,Dudek:2008sz}, e.g. $\sim \gamma_i (1 - \gamma_0)$ and $\gamma_5 (1 - \gamma_0)$, are much larger than the complementary combinations, e.g. $\sim \gamma_i (1 + \gamma_0)$ and $\gamma_5 (1 + \gamma_0)$, which would be zero in the non-relativistic limit.  This might be expected since the charmonium system is fairly non-relativistic.

For states within a given supermultiplet, it is expected that the $Z$-values for each of these operators, projected into the relevant lattice irreps, will be similar because they correspond to the same underlying spatial wavefunction with differing internal spin and angular momentum couplings.  This is seen to be the case in Fig.~\ref{fig:multiplets:conventional}, supporting our identification of these states as members of the P and D-wave supermultiplets.  We note that only qualitative conclusions can be drawn from the $Z$-values for lattice-regularised matrix elements; these require renormalisation to be compared with continuum matrix elements and this renormalisation can mix operators.

In summary, our extracted spectrum features states that follow the $n^{2S+1}L_J$ pattern expected by quark models along with others, having both exotic and non-exotic quantum numbers, that do not fit into this pattern.  We have argued that these extra states can be identified as hybrid mesons and discuss their phenomenology below after we compare our results with the experimental situation.

\subsection{Comparison with experimental results}
\label{sec:discussion:comparison}

It can be seen from Fig.~\ref{Fig:Summary} that the pattern of our extracted masses is in general agreement with the rather limited number of experimentally observed states.  We note that states above threshold can have large hadronic widths and a conservative approach is to only consider our mass values accurate up to the hadronic width~\cite{Dudek:2010wm}.  Discrepancies could be due to discretisation effects arising from the finite lattice spacing, systematic uncertainties within the variational method and because we have unphysically heavy up and down quarks.  We do not expect the unphysical up and down quarks masses to introduce a large error in the determination of lower-lying states~\cite{Bali:2011dc}.  However, a more careful treatment of these effects is needed in the vicinity of open-charm thresholds~\cite{Guo:2012tg}.  In addition, we have used the tree level value for the spatial clover coefficient and, as discussed in Section~\ref{subsec:c_sw}, this means that we underestimate the S-wave hyperfine splitting.  Further sources of systematic uncertainty arise from not including disconnected contributions, though these are expected to give only a small contribution in charmonium.  We have shown in Section \ref{sec:results:latticeirreps} that we do not see any finite volume effects at our level of statistical precision.

In the $1^{--}$ channel we identify the $\psi(2S)$ with our excited $^3S_1$ and the $\psi(3770)$ with our ground state $^3D_1$.  For these extracted states, we find that the admixture of $^3D_1$ in the dominantly $^3S_1$ state is small, as is the admixture of $^3S_1$ in the $^3D_1$ state.  We could not extract a significant signal for the $S-D$ mixing angle.  

Higher up in this channel there are two states in the PDG review, $\psi(4040)$ and $\psi(4160)$, whose robustness is not beyond question.  These states appear in the mass region where we have the second excitation of the $^3S_1$ with a relatively large uncertainty on the mass.  In light of the observations in Section \ref{subsec:nvecs}, a larger number of distillation vectors may be required in order to extract this state more precisely.  Alternatively, we may need to add operators with more radial structures to our basis, for example by including operators with additional spatial derivatives or different smearing profiles.

The $Y(4260)$ at $M-M_{\eta_c} \sim$ 1300~MeV with $J^{PC}=1^{--}$ is one of the states that appears to be supernumerary compared to the pattern expected by quark models.  A possible interpretation is as a non-exotic hybrid meson~\cite{Zhu:2005hp, Close:2005iz, Kou:2005gt}.  The mass of the $1^{--}$ hybrid candidate in our calculation agrees well with the mass of the $Y(4260)$, supporting the hybrid interpretation.  However, solely comparing masses does not allow strong conclusions to be drawn.  We find conventional $1^{--}$ charmonia in the same mass region and from our calculation we also can not exclude the possibility that the $Y(4260)$ is a multi-meson state or a tetraquark.

We find that the $1^{--}$ hybrid candidate has a significant overlap onto an operator which has the structure of a colour-octet quark-antiquark pair with $S=0$ in S-wave coupled to the gluonic field, $(\pi \times D^{[2]}_{J=1})^{J=1}$.  This is in contrast to the conventional $1^{--}$ mesons which have $S=1$ and is interesting because the $Y(4260)$ decays to $\pi^+ \pi^- J/\psi$~\cite{PDG2011}.  The folklore was that such decays should conserve the spin of the heavy quarks and so, because the $J/\psi$ has $S=1$, this would imply that the $Y(4260)$ also has $S=1$, at odds with our $S=0$ hybrid.  However, the observation by CLEO~\cite{CLEO:2011aa} of a cross section for $e^+ e^- \rightarrow \pi^+ \pi^- h_c$ (the $h_c$ has $S=0$) comparable to that of $e^+ e^- \rightarrow \pi^+ \pi^- J/\psi$ at a center of mass energy of 4170 MeV suggests that this folklore may not be correct.  Even so, the $Y(4260)$ being produced in $e^+ e^-$ annihilation suggests that it has some $^3S_1$ admixture.  In order to test the interpretation of the $Y(4260)$ and draw firmer conclusions about the nature of these vector mesons, the unsmeared decay constants $\sim \left<1^{--} \left| \bar{\psi} \gamma_i \psi \right | 0 \right>$ could be calculated for all the vector states we extract; this is beyond the scope of the present work.  In Section \ref{sec:discussion:hybrids} we discuss the more general hybrid phenomenology suggested by our results.

The $X(3872)$ at $M-M_{\eta_c} \approx$ 890~MeV, very close to $D \bar{D}^*$ threshold, is the `unexpected' state that has been confirmed by the largest number of experiments and perhaps is one of the most enigmatic; its structure remains unclear and is the subject of much discussion.  Possible $J^{PC}$ values for the $X(3872)$ are $2^{-+}$ and $1^{++}$.  In our results, the D-wave $2^{-+}$ state has $M-M_{\eta_c}$ around 30~MeV below the $X(3872)$, while the first excitation of the P-wave $1^{++}$ is around 110~MeV above the $X(3872)$.  Another possible interpretation for the $X(3872)$ is as a molecular state of $D$ and $D^*$ mesons.  As we have argued, we do not appear to see multi-hadron states in our extracted spectrum and so we would not expect to observe such a molecular state in this calculation.  To pin down the nature of the $X(3872)$ we plan to supplement our basis with operators constructed to overlap well with multi-meson states~\cite{Dudek:2010ew,Dudek:2012gj}.

\subsection{Exotic mesons and hybrid phenomenology}
\label{sec:discussion:hybrids}

\begin{figure}[t]
\begin{center}
\includegraphics[width=0.7\textwidth]{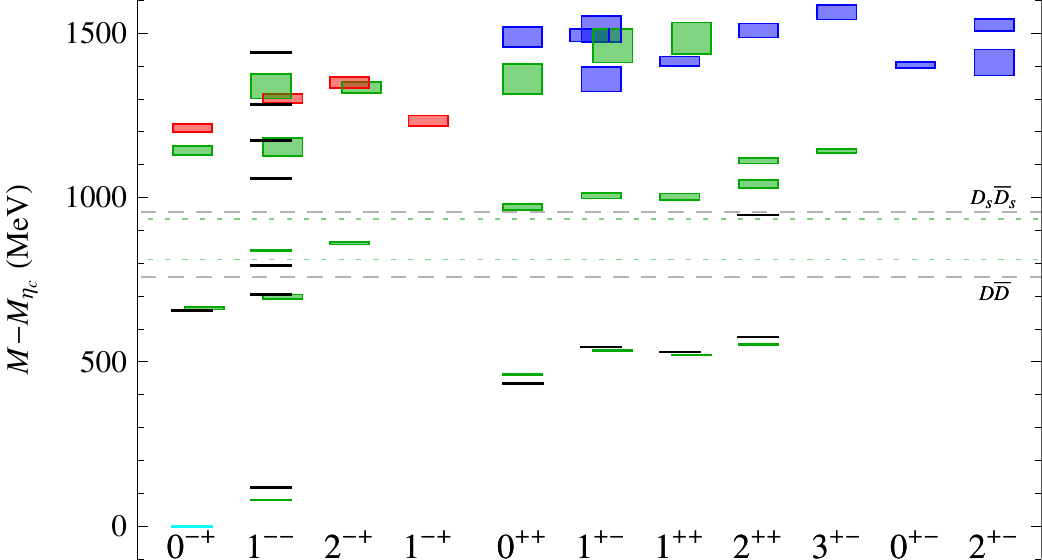}
\caption{Charmonium spectrum up to around $4.5~\text{GeV}$ showing only $J^{PC}$ channels in which we identify candidates for hybrid mesons.  Red (dark blue) boxes are states suggested to be members of the lightest (first excited) hybrid supermultiplet as described in the text and green boxes are other states, all calculated on the $24^3$ volume.  As in Fig.~\ref{Fig:Summary}, black lines are experimental values and the dashed lines indicate the lowest non-interacting $D\bar{D}$ and $D_s\bar{D}_s$ levels.}
\label{Fig:Hybrids}
\end{center}
\end{figure}

In Fig.~\ref{Fig:Hybrids} we show the charmonium spectrum for the subset of $J^{PC}$ channels in which, by considering operator-state overlaps, we identify candidate hybrid mesons.  A state is suggested to be dominantly hybrid in character if it has a relatively large overlap onto an operator proportional to the commutator of two covariant derivatives, the field-strength tensor.  We note that within QCD non-exotic hybrids can mix with conventional charmonia.  We find that the lightest exotic meson has $J^{PC}=1^{-+}$ and is nearly degenerate with the three states observed in the negative parity sector suggested to be non-exotic hybrids, $(0, 2)^{-+}, 1^{--}$.  Higher in mass there is a pair of states, $(0, 2)^{+-}$, and a second $2^{+-}$ state slightly above this.  Not shown on the figures, an excited $1^{-+}$ appears at around 4.6~GeV, there is an exotic $3^{-+}$ state at around 4.8~GeV and the lightest $0^{--}$ exotic does not appear until above 5~GeV.

The observation that there are four hybrid candidates nearly degenerate with $J^{PC} = (0, 1, 2)^{-+}, 1^{--}$, coloured red in Fig.~\ref{Fig:Hybrids}, is interesting.  This is the pattern of states predicted to form the lightest hybrid supermultiplet in the bag model~\cite{Barnes:1982tx, Chanowitz:1982qj} and the P-wave quasiparticle gluon approach~\cite{Guo:2008yz}, or more generally where a quark-antiquark pair in S-wave is coupled to a $1^{+-}$ chromomagnetic gluonic excitation as shown Table~\ref{table:multiplets}.  This is not the pattern expected in the flux-tube model~\cite{Isgur:1984bm} or with an S-wave quasigluon.  In addition, the observation of two $2^{+-}$ states, with one only slightly heavier than the other, appears to rule out the flux-tube model which does not predict two such states so close in mass.  The pattern of $J^{PC}$ of the lightest hybrids is the same as that observed in light meson sector~\cite{Dudek:2010wm,Dudek:2011bn}.  They appear at a mass scale of $1.2-1.3~\text{GeV}$ above the lightest conventional charmonia.  This suggests that the energy difference between the first gluonic excitation and the ground state in charmonium is comparable to that in the light meson~\cite{Dudek:2011bn} and baryon~\cite{Dudek:2012ag} sectors.

\begin{figure}[t]
\begin{center}
\includegraphics[width=0.6\textwidth]{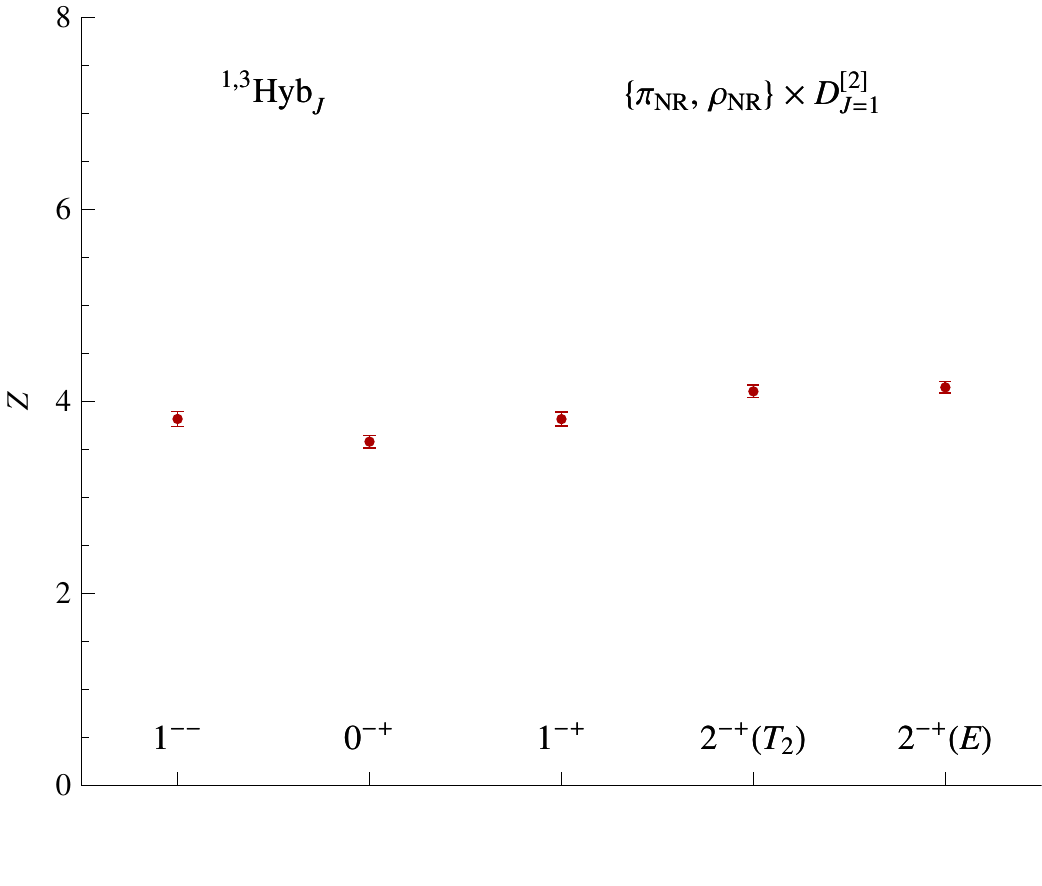}
\caption{Overlaps, $Z$, for the proposed lightest hybrid supermultiplet with operators $\{\pi,\rho\}_{\text{NR}}\times D_{J=1}^{[2]}$.}
\label{fig:multiplets:hybrid}
\end{center}
\end{figure}

To explore this hypothesis of a lightest hybrid multiplet further, we follow Ref.~\cite{Dudek:2011bn} and consider in more detail operator-state overlaps.  The operators $(\rho_{\text{NR}} \times D^{[2]}_{J=1})^{J=0, 1, 2}$ with $J^{PC} = (0, 1, 2)^{-+}$ and $(\pi_{\text{NR}} \times D^{[2]}_{J=1})^{J=1}$ with $J^{PC} = 1^{--}$ are discussed in that reference.  These operators have the structure of colour-octet quark-antiquark pair in S-wave with $S=1$ ($\rho_{\text{NR}}$) or $S=0$ ($\pi_{\text{NR}}$), coupled to a non-trivial chromomagnetic gluonic field with $J^{P_g C_g}_g = 1^{+-}$ where $J_g$, $P_g$ and $C_g$ refer to the quantum numbers of gluonic excitation.
Fig.~\ref{fig:multiplets:hybrid} shows that the four states suggested to form the lightest hybrid supermultiplet have considerable overlap onto operators with this structure.

For states within a given supermultiplet, it is expected that the $Z$-values for each of these operators, projected into the relevant lattice irreps, will be similar as discussed above.  The relevant overlaps presented in Fig.~\ref{fig:multiplets:hybrid} suggest that the four hybrid candidates have a common structure, supporting our identification of them as the members of the lightest hybrid supermultiplet consisting of S-wave $q\bar{q}$ coupled to a $1^{+-}$ gluonic excitation.  We note again that only qualitative conclusions can be drawn from the $Z$-values for lattice-regularised matrix elements.

A heavier hybrid supermultiplet composed of a P-wave colour-octet quark-antiquark pair coupled to a gluonic field with $J^{P_g C_g}_g = 1^{+-}$ would contain states with $J^{PC} = 0^{+-}$, $(1^{+-})^3$, $(2^{+-})^2$, $3^{+-}$, $0^{++}$, $1^{++}$, $2^{++}$ as shown in Table~\ref{table:multiplets}.  In Ref.~\cite{Dudek:2011bn} this was identified as the first excited hybrid supermultiplet in the light meson sector.  The exotic $0^{+-}$ state and two $2^{+-}$ states are observed in our spectrum with small mass splittings and these have relatively large overlap onto operators with the structure of a P-wave $q\bar{q}$ coupled to a gluonic $1^{+-}$.  In the non-exotic positive-parity sector, we have evidence from similar operator overlaps that in the region around $M-M_{\eta_c} \sim$ 1400 to 1500~MeV there are $0^{++}$, $1^{++}$, $2^{++}$ and $3^{+-}$ hybrids as well as candidates for three $1^{+-}$ hybrids.  We therefore observe candidates for all expected members of this excited hybrid supermultiplet and these are coloured dark blue in Fig.~\ref{Fig:Hybrids}.  To pin these states down more precisely we would need to add operators with more derivatives and spatial structures to our basis; this is beyond the scope of the current work.

In summary, we have identified possible lightest and first excited hybrid supermultiplets and suggest that these can be interpreted in terms of $q\bar{q}$ pairs in, respectively, S-wave and P-wave, coupled to a chromomagnetic gluonic excitation with $J^{P_g C_g}_g = 1^{+-}$.  The $0^{--}$ appearing at a much higher mass scale (above $\sim$ 5~GeV) may suggest a different type of gluonic excitation, for example a $1^{--}$ S-wave quasigluon, or have some other origin.  We note that the inclusion of multi-hadron operators in our basis could modify the interpretation of states above open-charm threshold.

\section{Summary}
\label{sec:summary}

Using distillation and the variational method with a large basis of carefully constructed operators, we have successfully computed an extensive spectrum of charmonium mesons with high statistical precision and reliably identified the continuum $J^{PC}$ of the extracted states. The large number of extracted states up to $4.5~\text{GeV}$, across all $PC$ combinations, goes beyond any other dynamical calculation.
The spin identification methodology has allowed us to identify the continuum spin of the extracted states and this will be crucial in order to compare the higher-lying states against observations once this dense spectrum has been more fully mapped out by experiments.

We find no statistically significant volume-dependence and also that the spectrum is stable with respect to changes in the details of the variational analysis and variation of the number of distillation vectors. 
Lattice discretisation effects were investigated by changing the spatial clover coefficient; however, a more complete determination of this systematic uncertainty would require additional lattice spacings.
The results presented here are for unphysically-heavy up and down quarks corresponding to a pion mass of 
$\approx 400~\text{MeV}$ at a single lattice spacing; lattice ensembles with lighter up and down quarks and further lattice 
spacings will be considered in future work. In our calculations we include only the connected quark line contributions 
to charmonium correlators. The disconnected 
contributions are expected to be small in charmonium since they are OZI suppressed. Nevertheless, in the distillation 
framework we can efficiently compute these contributions and determine their effect on the spectrum. 
This has already been done in the isoscalar light meson spectrum~\cite{Dudek:2011tt} and by applying the same methodology to 
charmonium we can determine the mixings between 
$c\bar{c}$, $s\bar{s}$ and $\tfrac{1}{\sqrt{2}}\left(u\bar{u} + d\bar{d}\right)$ basis states; these 
calculations are ongoing~\cite{Liu:2011rn}.

For the first time, we have computed the dynamical spectrum of charmonia with exotic quantum numbers 
($0^{+-}$, $1^{-+}$, $2^{+-}$) up to 4.5 GeV. A determination of these exotic states is particularly interesting because it points to the presence of explicit gluonic degrees of freedom. 
The spectrum of non-exotic states has many features in common with the $n^{2S+1}L_J$ pattern expected by quark models along with some states which do not appear to fit into such a classification. We have suggested that these extra states can be interpreted as non-exotic hybrids and have identified the lightest hybrid supermultiplet consisting of states with $J^{PC} = (0,1,2)^{-+},1^{--}$, as well as an excited hybrid supermultiplet. This is consistent with the pattern observed in lattice calculations in the light meson sector~\cite{Dudek:2011bn} and can be interpreted as a colour-octet quark-antiquark pair coupled to a $1^{+-}$ chromomagnetic gluonic excitation.
We find that the lightest gluonic excitation has an energy scale of $1.2-1.3~\text{GeV}$, comparable to that found in the light meson~\cite{Dudek:2011bn} and baryon~\cite{Dudek:2012ag} sectors, suggesting common physics.
Our results allow an interpretation of the $Y(4260)$ as a non-exotic vector hybrid meson, but based solely on comparing masses we can not draw a definitive conclusion; we have suggested further work to probe its structure.

We have presented arguments that, as in the study of light isovector mesons~\cite{Dudek:2010wm}, we see no clear evidence for multi-hadron states. 
To study such states we plan to enlarge the basis of operators to include those with a larger number of fermion fields. In 
Refs.~\cite{Dudek:2010ew,Dudek:2012gj} multi-hadron constructions have been presented and the method of L\"{u}scher and its extensions~\cite{Luscher:1991cf,Rummukainen:1995vs,Kim:2005gf,Christ:2005gi} was used to compute phase shifts 
from the denser spectrum of energy levels extracted.  The efficacy of these constructions was shown in application to $\pi\pi$ scattering in isospin-2. 
By applying this methodology to the charmonium sector we can determine the spectrum of multi-hadron states and then use L\"{u}scher's method and its 
extensions to compute phase shifts, and so determine the mass and width of resonances.

As well as the renaissance in charmonium spectroscopy, in recent years there has also been renewed interest in the spectroscopy of 
open-charm $D$ and $D_s$ mesons with unexpected states being observed. 
The methodology presented in this article can be applied to the study of these open-charm states and this work is in progress. 

An important goal of the Hadron Spectrum Collaboration is a full understanding of charmonium spectroscopy including properties of resonances. Experiments such as BESIII, PANDA and those at the LHC will collect vast numbers of $c\bar{c}$ states to explore the charm spectrum above and below threshold, including states with intrinsic gluonic excitations.  Our successful application of novel methods including distillation and spin-identification to determine the single-particle spectrum of charmonium up to $4.5$ GeV is a timely contribution to this effort.

\begin{acknowledgments}
We thank our colleagues within the Hadron Spectrum Collaboration.  
{\tt Chroma}~\cite{Edwards:2004sx} and {\tt QUDA}~\cite{Clark:2009wm,Babich:2010mu} were used to perform 
this work on the Lonsdale cluster maintained by the Trinity Centre for High Performance Computing 
funded through grants from Science Foundation Ireland (SFI), 
at the SFI/HEA Irish Centre for High-End Computing (ICHEC), 
and at Jefferson Laboratory under the USQCD Initiative and the LQCD ARRA project.  
Gauge configurations were generated using resources awarded from the U.S. Department of Energy 
INCITE program at the Oak Ridge Leadership Computing Facility at Oak Ridge National Laboratory, the NSF Teragrid at the Texas Advanced Computer Center 
and the Pittsburgh Supercomputer Center, as well as at Jefferson Lab.  
This research was supported by the European Union under Grant Agreement 
number 238353 (ITN STRONGnet) and by the Science Foundation Ireland under Grant Nos.~RFP-PHY-3201 and RFP-PHY-3218.
CET acknowledges support from a Marie Curie International Incoming Fellowship, 
PIIF-GA-2010-273320, within the 7th European Community Framework Programme. 
JJD, RGE, BJ and DGR acknowledge support from U.S. Department of Energy contract DE-AC05-06OR23177, under which Jefferson Science Associates, LLC, manages and operates Jefferson Laboratory. 
JJD also acknowledges the support of the Jeffress Memorial Fund and the U.S. Department of Energy Early Career award contract DE-SC0006765.
\end{acknowledgments}

\appendix
\section{Table of results}
\label{app:resultstable}

In Table~\ref{table:results} we tabulate the masses presented in Fig.~\ref{Fig:Summary}.  The calculated mass of the $\eta_c$ has been subtracted in order to reduce the systematic error from the tuning of the bare charm quark mass (see Section \ref{sec:lattices}). 

\begin{table*}[tb]
\begin{center}
\begin{tabular}{c|llllll}
$J^{PC}$ &\multicolumn{6}{c}{$(M-M_{\eta_c})$ / MeV}\\
 \hline
$0^{-+}$ & 0 & 663(3) & 1143(13) &1211(13) & & \\
$1^{--}$ &80.2(1) &698(6) & 840(3)   &1154(28) &1301(14) &1339(38) \\
$2^{-+}$ &860(3) &1334(17) &1350(17) & & & \\
$2^{--}$ &859(5) &1333(18) & & & &\\
$3^{--}$ &867(3) &1269(26) &1392(12)  & & &\\
$4^{-+}$ &1444(10) & & & & &\\
$4^{--}$ &1427(9) & & & & & \\
\hline
$0^{++}$ &461.6(7) &972(9) &1361(46) &1488(30)  & & \\
$1^{+-}$ &534(1) &1006(9) &1360(38) &1462(51) &1493(19) &1513(39)  \\
$1^{++}$ &521.6(9) &1002(10) &1415(14) &1484(48) & &\\
$2^{++}$ &554(1) &1041(12) &1112(8)  &1508(21) & & \\
$3^{+-}$ &1142(6) &1564(22) & & & &\\
$3^{++}$ &1130(9) & & & & & \\
$4^{++}$ &1129(9) & & & & & \\
\hline
$1^{-+}$ &1233(16) & & & & &\\
$0^{+-}$ &1402(9) & & & & & \\
$2^{+-}$ &1411(40) & 1525(18) & & & & \\
\end{tabular}
\caption{Summary of the charmonium spectrum calculated on the $24^3$ ensemble, as presented in Fig.~\ref{Fig:Summary}, with statistical uncertainties shown.  The scale has been set using the $\Omega$-baryon mass as described in Section~\ref{sec:lattices}; we note that $m_{\pi} \approx 400$ MeV in these simulations.  The calculated $\eta_c$ mass has been subtracted in order to reduce the systematic error from the tuning of the bare charm quark mass.  The masses of states with $J \ge 2$ are from joint fits to principal correlators as described in Section \ref{sec:spin}.} 
\label{table:results}
\end{center}
\end{table*}

\bibliography{paper}

\end{document}